\pdfoutput=1
\documentclass[pra,twocolumn,tightenlines,superscriptaddress,amsmath,amssymbs]{revtex4-1}
\usepackage[T1]{fontenc}
\usepackage[bbgreekl]{mathbbol}
\usepackage[normalem]{ulem}
\usepackage{graphics,graphicx,amsthm,amsmath,amssymb}
\usepackage{color}
\usepackage{mathtools}
\usepackage{setspace}
\usepackage{mathrsfs} 
\usepackage{float}
\usepackage[bottom]{footmisc}
\usepackage[caption = false]{subfig}
\usepackage{bm}
\usepackage{stackengine}
\usepackage{url}
\usepackage[english]{babel}
\usepackage{wasysym}
\usepackage{grffile}

\DeclareSymbolFontAlphabet{\mathbbm}{bbold}
\DeclareSymbolFontAlphabet{\mathbb}{AMSb}

\def\({\left(}
\def\){\right)}

\def\bra#1{\mathinner{\langle{#1}|}}
\def\ket#1{\mathinner{|{#1}\rangle}}
\def\braket#1#2{\mathinner{\langle{#1}|#2 \rangle}}

\def\exp{{\rm exp}}

\def\pt{t_{\rm PT }}
\def\B{{B_\perp}}

\def\e{{\epsilon}}

\def\scSc{\overline{\scS}}

\def\cA{{\mathcal A}}

\def\cO{{\mathcal O}}
\def\cE{{\mathcal E}}

\def\scP{{\mathscr P}}

\def\scH{{\mathscr H}}

\def\scS{{\mathscr S}}

\DeclareMathAlphabet{\mathpzc}{OT1}{pzc}{m}{it}

\makeatletter
\def\l@subsection#1#2{}
\def\l@subsubsection#1#2{}
\makeatother

\begin{document}
\title{Intermittency of dynamical phases in a quantum spin glass}

\author{Vadim N. Smelyanskiy}
\affiliation{Google, Venice, CA 90291, USA}
\author{Kostyantyn Kechedzhi}
\affiliation{Google, Venice, CA 90291, USA}
\affiliation{QuAIL, NASA Ames Research Center, Moffett Field, California 94035, USA}
\affiliation{University Space Research Association, 615 National Ave, Mountain View, CA 94043 }
\author{Sergio Boixo}
\author{Hartmut Neven}
\affiliation{Google, Venice, CA 90291, USA}
\author{Boris Altshuler}
\affiliation{Physics Department, Columbia University, 538 West 120th Street, New York, New York 10027, USA}

\date{\today}

\begin{abstract}
Answering  the  question of existence of efficient quantum algorithms for NP-hard problems require 
deep theoretical understanding of the  properties  of the  low-energy eigenstates and long-time coherent dynamics in quantum spin glasses.  We discovered and described analytically the property of asymptotic  orthogonality resulting in a new type of structure in quantum spin glass. Its eigen-spectrum is split into the alternating sequence of  bands formed by  quantum states of two distinct types ($x$ and $z$).  Those of  $z$-type are  non-ergodic extended eigenstates (NEE) in the basis of $\{\sigma_z\}$ operators that inherit the structure of the classical spin glass with  exponentially long decay times of Edwards Anderson order parameter at any finite value of transverse field $\B$. Those of $x$-type  form narrow bands of NEEs that conserve the integer-valued $x$-magnetization. Quantum evolution within a given band of each type is  described by a Hamiltonian that belongs to either the ensemble  of Preferred Basis Levi matrices ($z$-type) or Gaussian Orthogonal ensemble ($x$-type). We characterize the non-equilibrium dynamics using fractal dimension $D$ that depends on energy density (temperature) and plays a role of thermodynamic potential: $D=0$ in MBL phase,  $0<D<1$ in NEE phase,  $D\rightarrow 1$ in ergodic phase in  infinite  temperature limit. MBL states coexist with NEEs in the same range of energies even at very large $\B$. Bands of NEE states  can be used for new  quantum search-like algorithms of population transfer in the low-energy part of spin-configuration space. Remarkably, the intermitted structure of the eigenspectrum emerges in quantum version of a statistically featureless Random Energy Model  and is expected to exist in a class of paractically important NP-hard  problems that unlike REM can be implemented on a computer  with polynomial resources.

\end{abstract}
\maketitle
%


The structure of the  low-energy eigenstates and long-time coherent dynamics in quantum spin glasses is yet to be understood.
Until now the main emphasis was  on the study of the thermodynamical quantities and equilibrium phase diagrams. Rapid advances in quantum computation in recent decades  brought a significant interest in understanding  the coherent dynamics related to the   quantum   algorithms for discreet optimization and search problems, such as Quantum Annealing \cite{kadowaki1998quantum,farhi2001quantum,brooke1999quantum,smelyanskiy2002dynamics,boixo_evidence_2014,knysh2016zero,boixo2016computational,denchev2016computational,albash2018adiabatic}, Quantum Approximate Optimization \cite{farhi2014quantum} and Population Transfer \cite{IB,BaldwinLaumannEnergyMatching2018}.

The connection  between the properties of classical frustrated disordered systems and the structure of the solution space of random discreet optimization problems    has been noted first by Fu and Anderson \cite{fu1987application}. 
Energy landscapes in these problems  are characterized by the presence of a  large number of  deep local  minima far away from each other in spin-configuration space. This structure is at the root of computational hardness of   optimization problems  that require exponential  resources on a classical computer  \cite{mezard2009information}.

In quantum computing applications the spin glasses dynamics is often described  by a 
 Hamiltonian
\begin{gather}
  H =H_{\rm cl}+H_D\;, \label{eq:H}\\
 H_{\rm cl}= \sum_s \cE(s) \ket s\! \bra s,\quad  H_D =- \B \sum_{k=0}^n \sigma_x^k\;.\nonumber
\end{gather}
Here  the first term  encodes a discreet optimization problem with an energy function
 $\cE(s)$ defined  over the set of $2^n$
 configurations of $n$ Ising spins   ${s=(s^1,s^2,\ldots,s^n),}$ where $ s^k=\pm1$. It is  diagonal in the  eigenbasis~$\{\ket{s}\}$ of $\{\sigma_z^k\}_{k=1}^{n}$ operators to be referred to as the computational basis. In the second  term $\B$ denotes the transverse field that  gives rise to tunneling between the  quantum states defined within  individual spin-glass valleys. 
 
To understand the performance of quantum algorithms in spin glass models one should note that the number of possible transitions from a given state  $\ket{s} $ grows  exponentially with the
number $d$ of spin flips during the transition. This rapid growth can partially cancel the rapid decrease of the tunneling matrix elements with $d$ and lead to the eigenspectrum~\cite{IB}  that is qualitatively different from the finite-dimensional single-particle systems. The ergodic and localized parts of the spectrum are separated by the new phase of  non-ergodic extended eigenstates (NEE)~\cite{Mirlin2016}. They are sparse superpositions of spin configurations corresponding to deep local minima  within narrow energy belt at the tail of the density of states. 
These states are known to form bands with a rich structure of many-body correlations, including multifractal features, in both configuration and energy spaces~\cite{altshuler2016nonergodic}.
This  potentially makes such states of a substantial interest to  quantum discreet optimization and machine learning~\cite{carleoTroyer2017,biamonte2017quantum}.

Until now the quantum dynamics associated with NEEs has been neither discovered  nor described in practically relevant and implementable spin glass models. This problem 
remains challenging due to the extreme complexity of the distribution of the tunneling matrix elements between the  local minima of the spin glass energy landscape. 
Computing this distribution  numerically  can only be done via direct diagonalization for  a relatively small number of  spins where finite size effects dominate. As will become apparent below it also cannot be obtained with perturbative methods such as Forward Scattering Approximation (FSA).

There exist a broad class of random discreet optimization problems with important practical applications such as Number Partitioning Problem,  \cite{mertens1998phase,bauke2004number},  Multiprocessor Scheduling \cite{bauke2003phase},  Knapsack \cite{merkle1978hiding}, Binary Quadratic Programming, etc., that deal with high precision integer numbers and can be mapped onto classical Ising spin models with $2$-spin interactions.  These models can be
 encoded in a quantum circuit with a parameter set $\scP$ polynomial in $n$.
Rigorous analysis shows~\cite{borgs2001phase,mertens2000random,BaukeFranzMertens} that the statistical properties of these models in the low energy part of the spectrum are identical to those of the  Random Energy Model (REM)~\cite{derrida1981random}. This model is also important in its own right as a solvable example of a classical spin glass that arises in the theory of structural glasses~\cite{PhysRevB.36.8552} and the error correcting codes in information theory~\cite{mezard2009information}.

In REM the energies   $\cE(s)$ are samples
from the zero-mean normal  distribution with standard deviation $\sigma \sqrt{n}$. The density of states equals
\begin{equation}
\rho(\cE(s))=\frac{2^n}{\sqrt{2\pi n\sigma^2}} \exp\left(-\frac{\cE^2(s)}{2n\sigma^2}\right)\;. 
\label{eq:pREM}
\end{equation}
The  aforementioned models share several crucial properties (as illustrated in SM): (i) the energies in the low part of the spectrum $|\cE(s)|=\cO(n)$,
and the corresponding spin configurations  are statistically independent from each other;
(ii) the  energy function $\cE(s)$ can  be written in such a form that low energy states satisfy  $0> \cE(s)=\cO(n)$, the  typical value of $|\cE(s)|$ and the width of its distribution are  $\cO(n^{1/2})$, and flipping one spin in a low energy configuration changes energy by $\cO(n)$. In that region all states are deep local minima separated by $\cO(n)$ spin flips.

Despite lack of statistical correlations in the above models the specific many-body structure of NEE is encoded by the same parameter set $\scP$ as the model instance itself. This can potentially be used for data classification in analogy with quantum Boltzmann machines~\cite{amin2016quantum,biamonte2017quantum}. 

These properties of the REM  give rise to the two types of thermodynamic phases for   its  transverse  field quantum version (QREM) with Hamiltonian (\ref{eq:H}). They are  separated by the transition line $B_{\perp}(\beta)=\beta\,{\rm arccosh}({1\over 2}\exp(\beta \sqrt{\log 2}))$ where $\beta$ is inverse temperature. In the first type  $\B>B_{\perp}(\beta)$ the  free energy of the system is that of the independent spins in transverse field with finite temperature-dependent magnetization along $x$ axis.   In the second type  $\B>B_{\perp}(\beta)$ the free energy equals to that of the  classical REM and $x$-magnetization is zero. A naive   inference of the properties of non-equilibrium dynamics from  the above picture suggests  that
for $\B > B_{\perp}(\beta)$ the initial state $\ket{s}$ corresponding to a spin configuration $s$ with low-lying energy 
$|\cE(s)|=\cO(n)$ quickly decays into the eigenstates of the system with finite values of $\langle S_x\rangle =\cO(n)$,  each dominated by the classical energies $\cE(s)=\cO(\sqrt{n})$.

However a detailed analytical solution of non-equilibrium dynamics of QREM (\ref{eq:H}) reveals a qualitatively different  picture:
in a broad range of  transverse fields 
 \begin{equation}
 n^{1/2} \gg \B \gg n^{-1/2}\;,\label{eq:B-range}
 \end{equation}
most of the  states $\ket{s}$ corresponding to  low-energy spin configurations are  either exponentially (in $n$) long-lived or localized.
This picture is a result of the remarkable  intermittency feature in the low energy part of the  eigenspectrum of $H$.   The eigenspectrum is partitioned into   the alternating sequence of  bands  of  two qualitatively different types, which will be called $x$-type and $z$-type.

 Each $x$-type "miniband"   is the result of the splitting of a  level   $-2 \B m$ of the transverse field Hamiltonian $H_D$ (\ref{eq:H}) characterized by the conserved  total spin projection on x-axis $S_x=m$ where  $m\in[-n/2,n/2]$. The level degeneracy  is $\binom{n}{m}$ and the width equals
{\color{black}{$\Gamma_m^x=\sqrt{\sigma^2 n \, 2^{-n-1}\binom{n}{m}}$}}.  
  This splitting is caused by the REM Hamiltonian term  $H_{\rm cl}$ in (\ref{eq:H}) that couples the eigenstates 
  $\ket{x(m)}$ of  $H_D$  with the energy  $-2 \B m$.    The eigenstates of $H$ forming the miniband can  be obtained
  by the diagonalization of the $\binom{n}{m} $ - dimensional random matrix $\bra{x(m)}H_{\rm cl}\ket{x^\prime(m)}$ from the Gaussian Orthogonal Ensemble.  
  The  matrix elements $\bra{x(m)}H_{\rm cl}\ket{x^\prime(m^\prime)}$ connecting the subspaces with different values of $m$  lead to the renormalization of the average energies  of the minibands $-2 \B m$.  Remarkably,  in the entire range (\ref{eq:B-range}) they cause only a negligible $o(n^{-1})$  violation of the conservation of the total spin $x$-projection. 
  
  \begin{figure*}
  \includegraphics[width=7.0in]{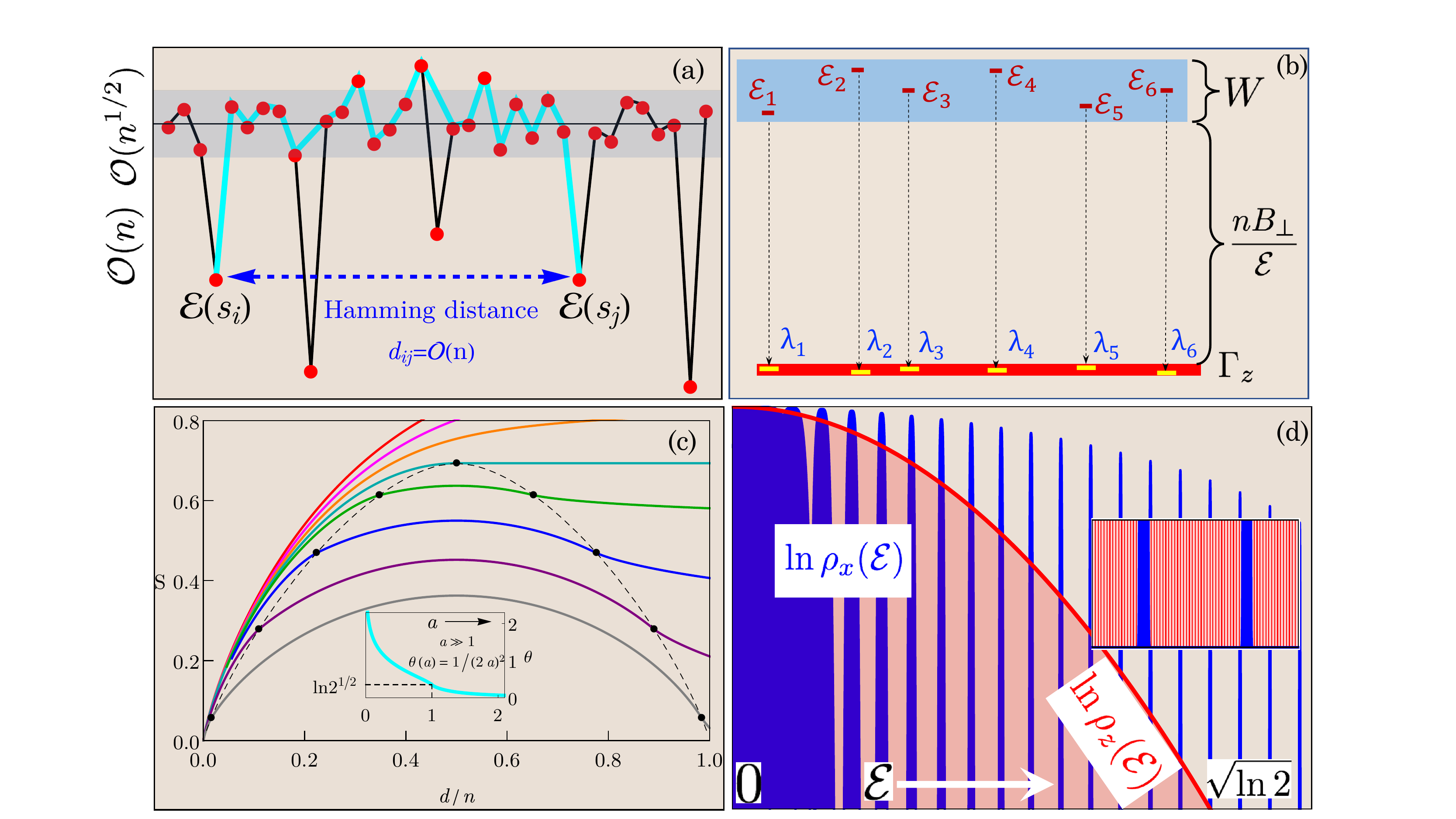}
  \caption{{\bf (a).}  Cartoon of the energy landscape of the binary optimization problem, $\cE(s)$ {\it vs} $s$, shown with red disks.  The horizontal axis corresponds to spin configurations $s$ and  vertical axis to $\cE(s)$. Dashed arrow indicate the large Hamming distance $d_{ij}=\cO(n)$ between the spin configurations $s_i$, $s_j$ with close  energies 
  $\cE(s_i),\cE(s_j)=\cO(n)$ connected by the tunneling paths that each proceed through the sequence of states separated by a single spin flip  in the central  band (gray color) with energies $\cE(s)=\cO(n^{1/2})$.  One such path is shown with broken cyan line. {\bf (b).} Carton depicts
 the  eigenvalues $\lambda_i$ (yellow lines) for NEEs forming an exponentially  (in $n$) narrow $z$-miniband $\Gamma^z$ (shown in red).
 The classical energies $\cE_i$ of the states $\ket{s_i}$ from the support set $\scS$ of the $z$-miniband  are shown with red lines, they belong to the energy strip  $W=\cO(1/n)$ shown with blue color.  The uniform shift of the energy strip relative to the $z$-miniband is $n\B/|\cE|=\cO(1)$.
  {\bf (c).}  Plots of the exponential factor in the matrix element (\ref{eq:Vd}) $S(\rho,a)=\lim_{n\rightarrow\infty}[V(n \rho,a)]^{1/n}$ {\it vs} rescaled Hamming distance $\rho=d/n$.  Different solid lines correspond to the values of $a$ from top to bottom: $a= $ 0.85(red), 0.9(magenta), 0.95(orange), 1.0(cyan), 1.05(green), 1.2(blue), 1.6(purple), 4(gray). Black points denote the values of $S(\rho,a)$ at the  boundaries $\rho_\pm=1/2\pm\mu_0$ of the region (\ref{eq:range}). Inside this region the $\rho$-dependence of $V$ is entirely due to the binomial factor in (\ref{eq:Vd}). The expression for $V(n\rho,a)$ outside this region is given in SM.  The inset is the plot of the exponential factor $\theta(a)$ in $V$ (\ref{eq:Vd}). {\bf  (d).} Cartoon of the log-plot of the density of  states $\ln\rho_z(|\cE|)$  {\it vs} $|\cE|$  for $z$-miniband (red) and $\ln\rho_x(|\cE|)$  {\it vs} $|\cE|$ for  $x$-miniband  (blue).  $\ln\rho_x(|\cE|)$ is peaked around  $|\cE|=\B |n-2m|,\, m\in[0,n]$ (eigenvalues of  $H_D$ (\ref{eq:H})) with peaks coalescing in the region
  $| \cE |=\cO(n^{1/2})$. Peak heights are decreasing with increasing  $|\cE|$. Thin red lines in inset depict $z$-minibands.   For $m>0$ the  width $\Gamma_m^x$ is at least $2^{-n/2}$ and the ratio $\Gamma_m^x/\Gamma^z(\cE)\gg 1$ is exponentially large in $n$  for $n/2+|\cE|/(2\B)\in[m, m+1]$. }
  \label{fig:landscape}
\end{figure*}

Eigenstates of $z$-type  compose the rest of the  low-energy spectrum outside of the exponentially small {\color{black}{ $\sim \Gamma_m^x$}} vicinities of the   energy values  $-2 \B m$.   For sufficiently low energies $|cE(s)|=\cO(n)$ the $z$-states are Many Body Localized (MBL) and their wave-functions are each  peaked at a single  local minimum and has a  weak admixture of the states from the middle of the spectrum $\cE(s)=\cO(n^{1/2})$.

As will be shown below,  above the MBL phase the $z$-eigenstates form NEE minibands via hybridization  of the exponentially many in $n$ computational basis states   $\ket{s}$ from narrow intervals  of the classical energies  with  the width $\cO(1/n)$.  These basis states are  separated by a large number  of spin flips,  their average $x$-magnetization is zero,  and the tunnelling matrix elements between them are  exponentially small  in $n$, which is at the root of their longevity as will be shown below.

Starting from a computational basis state  with energy $\cE$ the quantum evolution   is confined within the corresponding $z$-miniband  and  can be described by an effective downfolded Hamiltonian $\scH$
defined over a subset $\scS$ of  computational basis states $\ket{s}$ with low-lying energies inside the strip $ \left[\cE-W/2, \cE +W/2 \right]$, where $W\ll \B$.
The projection of the eigenstate $\ket{\Psi}$ of the full Hamiltonian  $H\ket{\Psi}=\lambda \ket{\Psi}$  onto the   low-energy subset  $\scS$ is a solution of the nonlinear eigenproblem  
\begin{gather}
\cE(s)\Psi_s+\sum_{s^\prime\in \scS} \Lambda_{s,s^\prime} (\lambda) \Psi_{s^\prime} =\lambda \Psi_s\;,\label{eq:neig}\\
\Lambda_{s,s^\prime}(\lambda)=\bra{s} H_D G(\lambda) H_D \ket{s^\prime}\,, \label{eq:Lambda}
\end{gather}
where $\Psi_s \equiv\braket{s}{\Psi}, \; s\in \scS$ and $G(\lambda)$ is a Green function  defined over the complement $\overline{\scS}$ of the subset $\scS$ in the Hilbert space so that 
\begin{gather}
G(\lambda)=(\lambda-H_0-P H_D P)^{-1},\label{eq:G}\\
H_0=\sum_{s\in \scSc} \cE(s)\ket{s}\bra{s}\;,\nonumber
\end{gather}
and $P=\sum_{s\in\scSc}\ket{s}\bra{s}$ is the projector onto  $\scSc$.

 We first   solve   Eq.~(\ref{eq:neig})  by neglecting the exponentially small  in $n$ off-diagonal  matrix elements $\Lambda_{s,s^\prime}(\lambda)$. The eigenvalues are given by  a series expansion in $1/n$ 
\begin{equation}
\lambda\approx \cE(s) +\frac{n B^2_{\perp}}{\cE}+\cO(1/n)\;.\label{eq:lambda0}
\end{equation}
Here the energy shifts depend quadratically  on  $\B$  because the  change in energy after a single spin flip $\cO(n)$ is much larger than transverse field $\B$ and disorder strength~$\cO(n^{1/2})$.  We choose the width $W$ to be much larger than the state-dependent dispersion of the  energy shifts  neglected in (\ref{eq:lambda0})  
\begin{gather}
W={\rm const}\times n^{-1}, \quad {\rm  const} \gg 1\;. 
 \label{eq:Wlimits}
\end{gather}
Including the   off-diagonal matrix elements in (\ref{eq:neig}) will give rise to  the repulsion between the eigenvalues (\ref{eq:lambda0}).

\textbf{ Off-diagonal matrix elements:}  The transverse field Hamiltonian $H_D=-\B\sum_{i=1}^{n}\sigma_{x}^{i}$  connects the states   separated by one spin flip.
Due to the statistical independence of the states in the low-energy subset $\scS$ the   typical Hamming distance between them (number of spin flips) is $d=n/2$ and the typical smallest distance is  extensive  $d=\cO(n)$. Therefore, the  effective  matrix element $\Lambda_{s,s^\prime}$ for the coupling between the two states    corresponds to the sum over  all elementary  spin-flip processes  with large number of spin flips. Each process  begins in the state $\ket{s}\in\scS$ and  reaches a high-energy state with typical energy $|\cE(s)|=\cO(\sqrt{n})$ in one spin flip, overcoming a large  energy gap  $\cO(n)$. It then proceeds through virtual high-energy states, and returns back to the subset $\scS$  only at the last step, at the state~$\ket{s^\prime}$, see Fig.~\ref{fig:landscape}(a).

Despite this deceptively simple picture,  the leading order term in the perturbation theory in $\B$ does not capture the behavior of  the off-diagonal matrix element $\Lambda_{s,s^\prime\neq s}$ even qualitatively in the regime   $n\B/|\lambda|\geq 1$.  There exists a cancellation between the  leading order and higher order in $\B$ terms  containing loops, i.e., multiple visits of the high-energy virtual states, see Fig.~\ref{fig:landscape}(d).

The long spin-flip paths in the above picture correspond to the tunneling processes with energy $\lambda$ connecting the deep  minima separated by an  energy barrier $\cO(n)$. Typical fluctuations of the barrier height $\cE(s)=\cO(\sqrt{n})$ are relatively small compared to its mean   and  determined  by the width of the classical  density function $\rho(\cE)$ (\ref{eq:pREM}), see Fig.~\ref{fig:landscape}(a).

The effect of the barrier fluctuations $\{\cE(s)\}_{s\in\scSc}$ on $\Lambda$, Eq.~(\ref{eq:Lambda}), can be inferred from the infinite series expansion of the Green function $G(\lambda)$ in terms of the disorder Hamiltonian $H_0$ given in~(\ref{eq:G}),
\begin{equation}
G(\lambda)=\sum_{k=0}^{\infty}G_0(\lambda)(H_0 G_0(\lambda))^k\;,\label{eq:dyson}
\;\; G_0(\lambda) = \frac{1}{\lambda-H_D}.
\end{equation}
Here $G(\lambda)$ is a random quantity determined by the instance of the disorder $H_0$. The bare Green function $G_0(\lambda)$ describes tunneling under the barrier without scattering off the height fluctuations. 

\begin{figure}
   \includegraphics[width= \columnwidth]{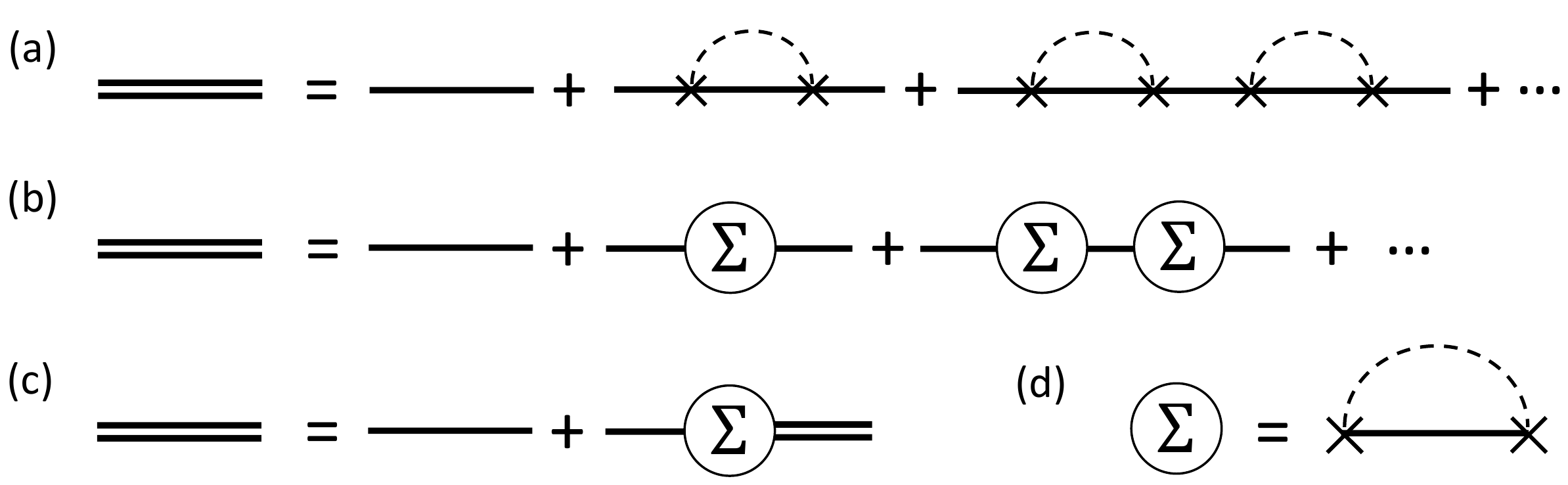}
  \caption{Diagrams 
    \label{fig:diagrams-main}}
\end{figure}

 We begin by computing the disorder average $\overline{ G}(\lambda)$  of the infinite series in the r.h.s of Eq.~(\ref{eq:dyson}) over the ensemble of $H_0$. Because $\{\cE(s)\}$ are  zero mean uncorrelated Gaussian random variables we use Wick's theorem to express the result in terms of the product of all possible pairings each corresponding to an element of the covariance matrix,
 \begin{gather}
 \overline{ \cE(s)\cE(s^\prime)}=\delta_{ss^\prime}n\sigma^2, \label{eq:Covariance}
 \end{gather}
 where $\delta_{ss^\prime}$ is the Kronecker delta. This gives the infinite series shown diagrammatically in Fig.~\ref{fig:diagrams-main}(a).
 
Applying the disordered diagrammatic technique ~\cite{AGD:1963} we introduce $\Sigma(\lambda)$ which is the sum of all irreducible diagrams  in the disorder-averaged series in  Fig.~\ref{fig:diagrams-main}(a). A diagram is irreducible if it cannot be split into disconnected parts by cutting a bare Green function $G_0^{-1}$. This results in the Dyson series shown in Fig.~\ref{fig:diagrams-main}(b)  that can be resumed to obtain the Dyson equation, 
\begin{gather}
\overline{G}(\lambda)= G_0(\lambda)+G_0(\lambda)\Sigma(\lambda) \overline{G}(\lambda).
\end{gather}
The structure of the diagrams in the sum in Fig.~\ref{fig:diagrams-main} is complicated since operators are defined in the $2^n$-dimensional Hilbert space. Nonetheless, we show that $\Sigma$ is well approximated by the  leading order diagram depicted in Fig.~\ref{fig:diagrams-main}(d) and proportional to an identity matrix $\hat{I}$, \textcolor{black}{so that the only effect of disorder is to renormalize the energy, $\lambda\rightarrow \bar{\lambda}$}
\begin{gather}
\Sigma(\lambda) = \frac{n\sigma^2}{2\cE} \hat{I}+\cO\left(\frac{1}{n}\right),\quad \bar{\lambda}=\lambda-\frac{n\sigma^2}{2\cE}
\end{gather} 
While the details are given in the SM, here we provide a qualitative picture explaining the result. 
Because the transverse field Hamiltonian $H_D$ is symmetric with respect to permutations of individual spins the  matrix elements of  $G_0(\lambda)$,
\begin{gather}
\bra{s} G_0(\lambda) \ket{s^\prime} =  G_0\left( \lambda,d_{ss^{\prime}}\right), \,\,
\bra{s} G_0(\lambda) \ket{s}\approx \lambda^{-1}\;, \label{eq:G0}
\end{gather}
depend only on the Hamming distance $d_{ss^{\prime}}$ between the spin configurations $s, s^\prime$, decaying exponentially with $d_{ss^{\prime}}$. As a result,  $G_0(\lambda,d)$ corresponds to an effectively one-dimensional tunneling under the barrier 
 $|\lambda| =\cO(n)$ that is well described by an eikonal approach Ref.~\onlinecite{IB}. 
The wave function under the barrier changes significantly when Hamming distance changes by $\delta d \sim 1$. On this scale the correction to the eikonal due to the disorder is $\cE(s)/\lambda$. Because of the high barrier the  disorder in Eq.~(\ref{eq:Covariance}) is effectively weak and the eikonal corrections are small, $\overline{\left(\cE(s)/\lambda\right)}^2 \sim 1/n$. Therefore the effective "scattering length" is $\cO(n)$ and much larger than the "de Broile wavelength" under the barrier which is $\cO(1)$. As a result the weak disorder perturbation theory is asymptotically correct in the limit $n\gg1$.

It follows from the above that the relative standard deviation (RSD) of the off-diagonal matrix elements $\mathrm{RSD}\left(\Lambda_{s,s^\prime}\right) = \cO(\sigma/n)$ is small.
Therefore $\Lambda_{s,s'\neq s} (\lambda)$ in the non-linear eigenvalue problem (\ref{eq:neig}) are well approximated by their disorder averaged values,
\begin{gather}
\Lambda_{s,s^\prime}(\lambda) \approx \overline{\Lambda}_{s,s^\prime}(\lambda)= \lambda^2 G_0\left(\bar{\lambda}, d_{ss^\prime}\right),
\end{gather}
($s\neq s^\prime$). It follows from (\ref{eq:dyson}) that $\overline{\Lambda}_{s,s^\prime\neq s}(\lambda)$ changes significantly on the scale of the level separation $2\B$ of the driver Hamiltonian $H_D$ that greatly exceeds  the strip width $W$ (\ref{eq:Wlimits}). Therefore  $\overline{\Lambda}_{s,s^\prime\neq s}(\lambda)\simeq \Lambda_{ss^\prime}(\cE)$, reducing the non-linear eigenproblem to a linear one.  

\textbf{Downfolded Hamiltonian:} Using the asymptotic expression for $G_0(\lambda, d)$ at  $d=\cO(n)$ \cite{IB}, the matrix elements of the linearized downfolded Hamiltonian  $\scH$ can be written as, 
\begin{gather}
\scH_{s,s^\prime}=\delta_{ss^\prime} \e_{s} + \left( 1-\delta_{ss^\prime}\right)  V(d_{ss'},a) \sqrt{2}\sin \phi_{ss^\prime}, 
\label{eq:scH}
\end{gather}
\begin{gather}
V(n\rho,a) =\cA(\rho,a) 
\frac{n^{5/4} e^{-n \theta(a)}}{\sqrt{\binom{n}{n\rho}}}. \label{eq:Vd}
\end{gather}
where  $\e_s=\cE(s)-\cE$ and we subtracted  a uniform shift  $\cE +n B^2_{\perp}/\cE$. The energy per spin $e=\cE/n$ and the expression (\ref{eq:Vd}) applies for sufficiently large  transverse fields  $\B$ and   Hamming distances  $d=n\rho$ 
\begin{gather}
a\equiv \B/|e|>1,\quad  |1-2d/n | \leq \mu_0=\sqrt{1-a^{-2}}\;.\label{eq:range}
\end{gather}
The phase $\phi_{ss^\prime}=\phi( d_{ss^\prime})$ in (\ref{eq:scH}) describes fast oscillations of tunneling matrix elements under the barrier with $d$. The amplitude  $V(d,a)$ (\ref{eq:Vd}) is shown in Fig.~\ref{fig:landscape}(d), its behavior in the range (\ref{eq:range}) is predominantly determined by the binomial coefficient, while the prefactor $\cA$ is a smooth function of $\rho$.  The function $\theta(a)$ is shown in the inset of Fig.~\ref{fig:landscape}(d),  and $\theta(a)\approx 1/(4a^2)$ for $a\gg 1$.

A total weight $Q=\sum_{s\in\scS}|\Psi_s|^2$ of a low energy  eigenstate in the  downfolding subspace  $\scS$ corresponding to  the energy strip $W$ can  be calculated  within the weak disorder perturbation theory,  $Q= -1/(\cE^{2}\partial_\lambda G_0(\lambda,0))+\cO(1/n)$.  Using Eq.~(\ref{eq:G0}) we obtain $Q=1-\B^2/n+\cO(1/n)$, where omitted $1/n$ corrections are due to disorder. This underscores a crucial observation  that under the condition (\ref{eq:B-range})
the dense Hamiltonian $\scH$ (\ref{eq:scH}) provides a self-contained description of the eigenspectrum of QREM within  the Hilbert subspace spanned by the $M=W\rho(\cE)$ basis vectors $\ket{s}$  with  energies $\cE(s)$ inside the narrow strip of the width $W$ centered at $\cE=n e$.

{\bf The eigenspectrum of ${\bm z}$-type:}
The eigenspectrum of the random matrix $\scH$ has been studied in a different context~\cite{IB}. Its  rescaled  off-diagonal matrix elements $x\equiv V(d_{ss^\prime},a)/V(n/2,a)$ obey the
 heavy-tailed distribution  
 $P(x)= c\,  x^3/\sqrt{\ln x}$   
 defined over the interval $x\geq 1$ 
with the maximum at $x=1$  corresponding to a typical Hamming distance  $d_{ss^\prime}=n/2$. The typical off-diagonal matrix element  
\begin{gather}
V(n/2,a) \propto 2^{-n/2} e^{-n\theta(a)}\;,\label{eq:Vtyp}
 \end{gather}
is much smaller than the  dispersion of the diagonal elements, $W\gg V(n/2)$.  The random matrix $\scH$ belongs to the ensemble of Preferred Basis Levi Matrices (PBLM)  \cite{IB}. 

\begin{figure}[t]
   \includegraphics[width= \columnwidth]{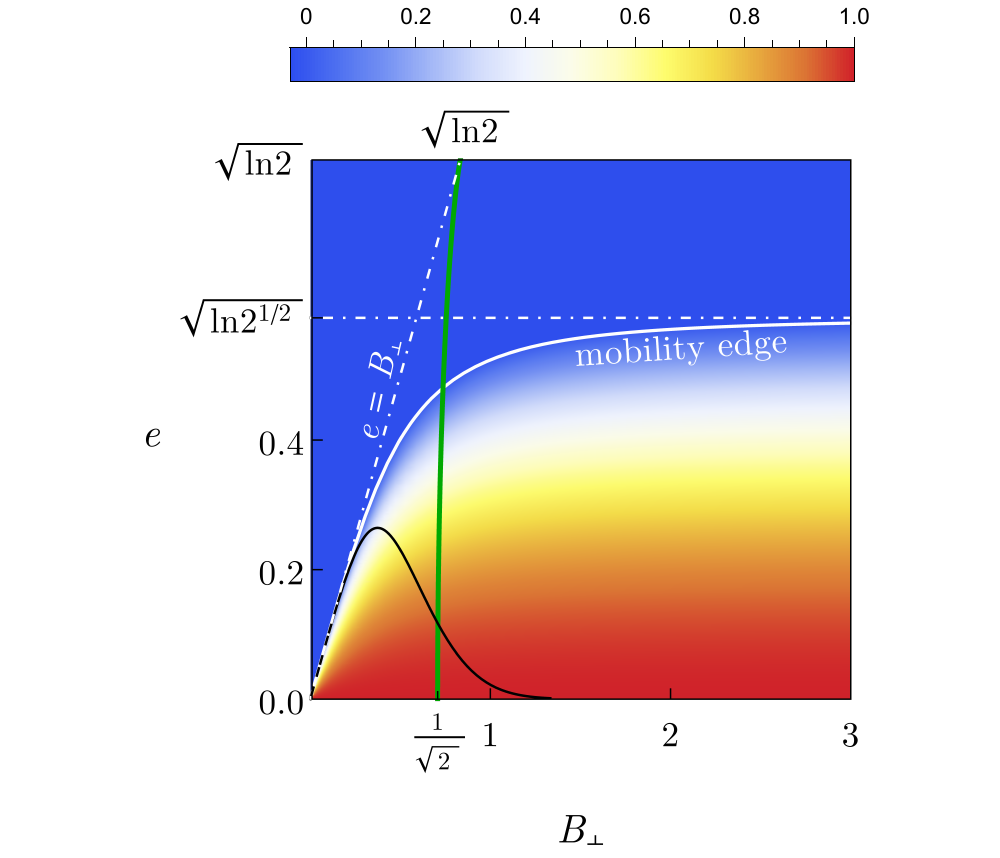}
  \caption{Density plot of the fractal dimension $D=D(|e|,\B)$ (\ref{eq:Omega}). {\color{black}{Maximum of}} $|e|=\sqrt{\ln 2}$ corresponds to  typical lowest (highest) energy density  in REM. Mobility edge, $D(|e|,\B)=0$, separating MBL and NEE phases is shown with white line. 
Dot-dashed lines show its the asymptotes. Blue area depicts MBL phase \textcolor{black}{existing}  at all finite values of $\B$.   In the area to the right of the "phase transition"  line  shown in green color the cumulative density of $z$-states is greater than that of the $x$-states, the opposite condition holds to the left of the line \textcolor{black}{(see SM)}. Area encompassed by the black line corresponds  to NEE states formed by the transitions to nearest resonances.  
    \label{fig:fractalD}}
\end{figure}

The control parameter of the ensemble  is the ratio of the typical off-diagonal matrix element to the mean inter-level spacing, $\gamma=V(n/2,a)\rho(n e)$,  determined by the value of  $e$ at the  energy strip and transverse field  $\B$.
 For $\gamma <1$ the eigenstates of $\scH$ are many-body localized. For $\gamma>1$   the eigenspectrum  of $\scH$  splits into a large number of minibands of NEEs. The eigenstates $\braket{s}{\psi}$ from a given miniband  are  peaked at the same states $\ket{s}$ forming the support set of the miniband $\scS$ that is  sparse in the computational basis. The size of the set $|\scS|=\Omega$ scales  exponentially with $n$,  yet it remains exponentially small compared to the size of the  full Hilbert  space $2^n$. Fractal dimension $D$ of the set $\scS$ equals (see also Fig.~\ref{fig:fractalD})
\begin{gather}
D=\lim_{n\rightarrow \infty}\frac{\log_2\Omega}{n}, \quad D= 1- \frac{2e^2}{\sigma \log 2} - \frac{2 \theta(\B/e)}{\log2}. \label{eq:Omega}
\end{gather}
 We note that in our spin glass model unlike NEEs in tight binding models on random graphs the fractal dimension   explicitly depends on energy density (or temperature). 

  For the set of eigenstates $\ket{\psi_\beta}$ in a given NEE miniband one can define a corresponding 
 spread of the eigenvalues $\lambda_\beta$, the miniband width $\Gamma^z$.  In the matrix ensemble of  $\scH$ the width   $\Gamma^z$  is a random variable that obeys a Levi stable distribution  ~\cite{IB} 
  {\color{black}{with   typical value $\Gamma^{z}_{\rm typ} \propto  V(n/2,a) 2^{n D/2}$   and characteristic dispersion $C$  that
depend on the fractal dimension $D(e,\B)$
\begin{equation}
\Gamma^{z}_{\rm typ}\sim C \propto 2^{-\frac{1-D}{2}}\e^{-n\theta}\;.\label{eq:Gamma-typ}
\end{equation}
Now the physical meaning of the  energy strip   $W\sim 1/n\gg \Gamma^{z}_{\rm typ}$ (\ref{eq:Wlimits})  becomes transparent:}}  it corresponds to the dispersion of classical energies $\cE(s)$ of the states $\ket{s}\in\scS$ from the support set of a {\it given} miniband, see Fig.~\ref{fig:landscape}(b).

{\color{black}{This structure of NEEs manifests in the non-equilibrium dynamics and  can be observed in the following experiment on quantum computer implementing "population transfer"  \cite{IB}  protocol.  Starting from an initial state $\ket{s_0}$ with classical  energy $\cE(s_0)$ 
we let the system evolve with the Hamiltonian $H$ at constant value of $\B$ over a time of population transfer $\pt \apprge 1/\Gamma^{z}_{\rm typ}$ where a finite fraction of the miniband support set $\scS$ will be popiulated.}} We then perform measurements in the computational basis. 
The dispersion of classical energies $\cE(s)$ increases from $0$ to $W$ over the time  scale $t \apprge  \pt$.
The distribution of miniband widths could be observed using phase estimation on a quantum computer or in a simpler setting by studying the statistics of the Edwards-Anderson (EA) order parameter ${ q_{EA}(t, s_0)= n^{-1}  \sum_{i=1}^n  \langle \sigma^i_z(t) \sigma_z^i(0) \rangle}$. At short times $ q_{EA}(t, s_0)\approx 1$ because spins are frozen parallel to $z$-axis.  On the time scale $t \apprge  \pt$ the order parameter decays as~$e^{-\Gamma^z(s_0) t}$ due to the tunneling from $\ket{s_0}$ into the states from the same miniband. Starting from different initial states $s$ distribution of minibands width $p(\Gamma^z)$ can be determined as a function of energy density $e$ and transverse field $\B$. At the longer time scale $t\gg \pt$, {\color{black}{ $q_{EA}(\infty, s_0)\sim  2^{-n D(e, \B)}$. }}

The above asymptotic value $q_{EA}(\infty,s_0)$ of the EA  order parameter undergoes a sharp  transition from exponentially small in $n$ value to $q_{EA}\approx 1$ at the mobility edge $e=e_c(\B)$, where the fractal dimension $D(e_c, \B)=0$, corresponding to the onset of MBL phase. The form of the mobility edge   is given in Fig.~\ref{fig:fractalD}. It is a nearly linear function $e_c \approx \B+2B^3_{\perp}/\log\B$  at small values of $\B\ll1$. At large values $\B \gg1$ it energy saturates as $e_c\approx \sqrt{\log\sqrt{2}}(1-1/(8B_{\perp}^2))$.

{\bf Decay Rates:} The exponential decay of a given state $\ket{s_0}$ is a sum of contributions from  a large number of channels. Each of them  corresponds to a decay with the typical  rate $\Gamma^{z}_{\rm typ}(d)$ into the subset $\Omega(d) = 2^{(1-D)n} \binom{n}{d}$  of   states  $\ket{s}$ located on a Hamming distance $d$ within the same miniband as the state  $\ket{s}$.  Remarkably,  the  decay rates   $\Gamma^{z}_{\rm typ}(d) \sim \Gamma^{z}_{\rm typ}/n$ do not depend on $d$ when it is sufficiently large   (\ref{eq:range}). However   the dominant transitions occur with  $d=n/2$ spin flips into the channel with the maximum number of states.  This  "equipartition" of decay rates happens because the steep decrease of the squared matrix element $V^2(d,a)$ (\ref{eq:Vd}) with $d$ is cancelled by the steep increase in the number of states $\Omega(d)$ in the $d$-charnel.

The closest state in a miniband to a given state $\ket{s_0}$ is located at the distance $d_{\rm min}$  where  $\Omega(d_{\rm min})=0$. 
{\color{black}{In a certain region of parameters $(\B,e)$ (area under black line in Fig.~\ref{fig:fractalD})
 the value of $d_{\rm min}$ is sufficiently small so that the condition (\ref{eq:range}) is violated. }} As can be seen from Fig.~\ref{fig:landscape}(d), in this case the decrease of  $V(d,a)$  with $d$ is exponentially  steeper than  that given in Eq.~(\ref{eq:Vd}).  Before a   state $\ket{s_0}$  has a chance to decay into the  states on a distance $d=n/2$ it will hybridize with its nearest neighbors, they will hybridize with their own nearest neighbors, etc,  forming a tree of "resonances" inside the miniband. 
The width of the miniband in this case is determined by the  matrix element to the nearest resonance $V(d_{\rm min})$ (see SM for details).  After a sufficiently long time the transition channels with $d=n/2$ will begin to play a role and eventually all states inside the band $V(d_{\rm min})$ will be populated. The fractal dimension of the eigenstates in the miniband is  $D=\lim_{n\rightarrow \infty} [V(d_{\rm min})\rho(n e)]^{1/n}$ where $e$ is the energy density at  the miniband (see  SM).

 {\bf Asymptotic orthogonality:} Remarkably,  MBL and  low-energy NEE phases exist  in the entire range  of $\B$  (\ref{eq:B-range}), even  at $\B\gg1$, see Fig.~\ref{fig:fractalD}. 
 This happens because for $\B\gg1$ the weight of the low-energy eigenstate $\ket{\Psi}$ with the eigenvalue $\lambda \simeq n e$ at a  typical off-resonant computational basis state with energy $\cE(s)=\cO(n^{1/2})$ is $\sim G_0(\lambda,n/2)^2 \propto 2^{-n}e^{-n e^2/(2B_{\perp}^{2})}$. Whereas the number of such states  is $\simeq 2^n$. Therefore the cumulative weight of  $\ket{\Psi}$ over the typical off-resonant states is exponentially small  $\sim e^{-n e^2/(2B_{\perp}^{2})}$. The cumulative weight of $\ket{\Psi}$ over  atypical states with  $\cE(s)=\cO(n^{1/2})$ (close in Hamming distance to the  support set of $\ket{\Psi}$) is also small, $\cO(1/n)$. This  "asymptotic orthogonality"  prevents  substantial mixing of the low energy support sets   and leads to the  longevity  of the   $z$- and $x$-states even at the  large transverse fields $\B\gg1$ (\ref{eq:B-range}).
 
This orthogonality feature is ubiquitous in a number of computational problems that share statistical features of low energy spectrum with REM described in the paper. This coexistence of bands of NEE and MBL states for any finite ($n$-independent) transverse field is a quintessentially quantum effect and a qualitatively new feature of the quantum spin glass models. This behavior is not limited to spin glasses in transverse field and could be generalized to other local driver Hamiltonians. The resulting non-equilibrium dynamics of NEE in the  quantum spin glass and the many-body correlations across their support sets may have important implications for quantum population transfer ~\cite{IB,BaldwinLaumannEnergyMatching2018},  reverse annealing~\cite{ReverseAnnealing} and quantum machine learning \cite{PhysRevX.8.021050,li2018quantum}.

\begin{acknowledgments}
K.K. acknowledges support by NASA Academic Mission Services, contract number NNA16BD14C.  This research is based upon work supported in part by the AFRL Information Directorate under grant F4HBKC4162G001 and the Office of the Director of National Intelligence (ODNI) and the Intelligence Advanced Research Projects Activity (IARPA), via IAA 145483. The views and conclusions contained herein are those of the authors and should not be interpreted as necessarily representing the official policies or endorsements, either expressed or implied, of ODNI, IARPA, AFRL, or the U.S. Government. The U.S. Government is authorized to reproduce and distribute reprints for Governmental  purpose notwithstanding any copyright annotation thereon.
\end{acknowledgments}.



%


\appendix
\section{Details of disorder diagrammatic calculations}

\subsection{Bare Green function at large Hamming distances $d=\cO(n)$ }

In this section we provide the details of the calculation of the matrix element of the Green function at large Hamming distance  $d=\cO(n)$. Eigenstates of QREM in the finite energy density subspace $\scS$ can be found from the non-linear eigenvalue equation, Eq.~(\ref{eq:neig}), which contains the Green function $G(\lambda)$ of the subspace $\overline{\scS}$ taken at the low energy $\left|\lambda\right|\sim \mathcal{O}(n)$. $G(\lambda)$ is the inverse of a sum of two non-commuting operators: $\lambda -H_0$ diagonal in the computation basis and $H_D = - 2 B_\perp \hat{S}^x$ as a results its the explicit expression is not trivial. Perturbation theory in weak transverse field $B_{\perp}/\epsilon \ll 1$ gives adequate qualitative description of the eignestates within the many-body localized region. In this regime, each eigenstate has a sharp peak at a single bitstring weakly dressed by coupling to bitstrings at Hamming distances $d\sim \mathcal{O}(n^0)$.  Outside  many-body localized regime REM eigenstates are qualitatively different and their description requires careful non-perturbative calculation of $G(\lambda)$. The first step in such calculation is to consider matrix elements of the disorder-free Green function in the basis of bistrings $\ket{z^i}\equiv \otimes_{k=1}^n  \ket{z^i_k}$,
\begin{gather}
G(\lambda, d_{ij})\equiv  \bra{z^j}(\lambda +2  \hat{S}^x B_{\perp} )^{-1}\ket{z^i} \nonumber\\
=
 \sum_{l=1}^{2^n} 
\frac{
\braket{z_j}{x^l}\braket{x^l}{z_i} }{\lambda+2 B_\perp  \sum_{k=1}^n x_k^l 
}. \label{eq:Gdij}
\end{gather}
Denominator in the sum over $\ket{x^l}\equiv \otimes_{k=1}^{n}  \ket{x_k^l}$  in Eq.~(\ref{eq:Gdij}) depends only on the total magnetization along $x$-axis $\hat{S}^x\ket{x^l}=\sum_{k} x_k^l \ket{x^l}$. Therefore $G(\lambda, d_{ij})$ is symmetric with respect to permutations of bits in bitstring $z^i\oplus z^j$, and depends only on the Hamming distance $d_{i,j}$. Using this symmetry Eq.~(\ref{eq:Gdij}) can be rewritten as a sum over Hamming distances $d$,
\begin{gather}
 c(a, d)=  \sum_{k=0}^d\sum_{p=0}^{n-d} \left(
\begin{array}{c}
 d \\
 k \\
\end{array}
\right) \left(
\begin{array}{c}
 n-d \\
 p \\
\end{array}
\right)\frac{(-1)^k 2^{-n}}{1-\frac{n}{a\left(n-2 (k+p)\right)}},\label{eq:CExact}
\end{gather}
where $ G(a, d)\equiv \left( \delta(d) -c(a, d)\right) \lambda^{-1} $, and $a\equiv n B_\perp/\left|\lambda\right|$.

We further exploit the permutation symmetry of $G(\lambda, d_{ij})$ by setting $\ket{z^{i=0}}=\ket{0...0}$, which corresponds to trivial relabeling of bitstrings. In this case the Green function is proportional to the projector on the permutation symmetric subspace,
\begin{gather}
\hat{\mathcal{P}}_{n/2} \equiv \sum_{m} \ket{m,\frac{n}{2}} \bra{m,\frac{n}{2}},
\end{gather}
where $\sum_{\alpha=x,y,z} \left(\hat{S}^\alpha \right)^2 \ket{m,\frac{n}{2}}=\sqrt{\frac{n}{2}\left(\frac{n}{2}+1\right)}\ket{m,\frac{n}{2}}$ and $\hat{S}^z\ket{m,\frac{n}{2}}=\left(\frac{n}{2}-m\right)\ket{m,\frac{n}{2}}$,
\begin{gather}
G(\lambda,d)=
 \bra{z^j}
\hat{\mathcal{P}}_{n/2}
 (\lambda +2  \hat{S}^x B_{\perp} )^{-1}
 \ket{z^0} 
=
 \frac{\mathbb{G}_{\frac{n}{2}-d,\frac{n}{2}}}{\sqrt{\binom{n}{d}
}},  \label{eq:Gbold}
\end{gather}
Here we related Hamming distance to $z$-axis magnetization $m=\frac{n}{2}-d$ and used, 
\begin{gather}
\braket{z^j}{m,\frac{n}{2}} =\frac{\delta_{m,\sum_k z_k^j}}{\sqrt{\binom{n}{n-2m}}}.
\end{gather}
In Eq.~(\ref{eq:Gbold})  we recognize the Green function of a large spin $\mathbb{G}_{\frac{n}{2}-d,\frac{n}{2}}$ in the maximum spin subspace. It satisfies a tridiagonal recurrence relation, 
\begin{gather}
\delta_{m,\frac{n}{2}} -\sum_{\alpha=\pm} u\left(m+\frac{\alpha}{2}\right) \mathbb{G}_{m-\alpha,\frac{n}{2}} 
=
\lambda \mathbb{G}_{m,\frac{n}{2}}, \label{eq:recur}\\
u(m)=B_\perp \sqrt{L^2-m^2},\;\;\; L\equiv \frac{n+1}{2},
\end{gather}
where we approximated $\left(\hat{S}^x\pm i\hat{S}^y\right) \ket{m,\frac{n}{2}}\approx \sqrt{L^2-(m\pm\frac{1}{2})^2}\ket{m\pm1,\frac{n}{2}}+\mathcal{O}(\frac{1}{n} )$.

\begin{figure}[h]
\includegraphics[width=0.99\columnwidth]{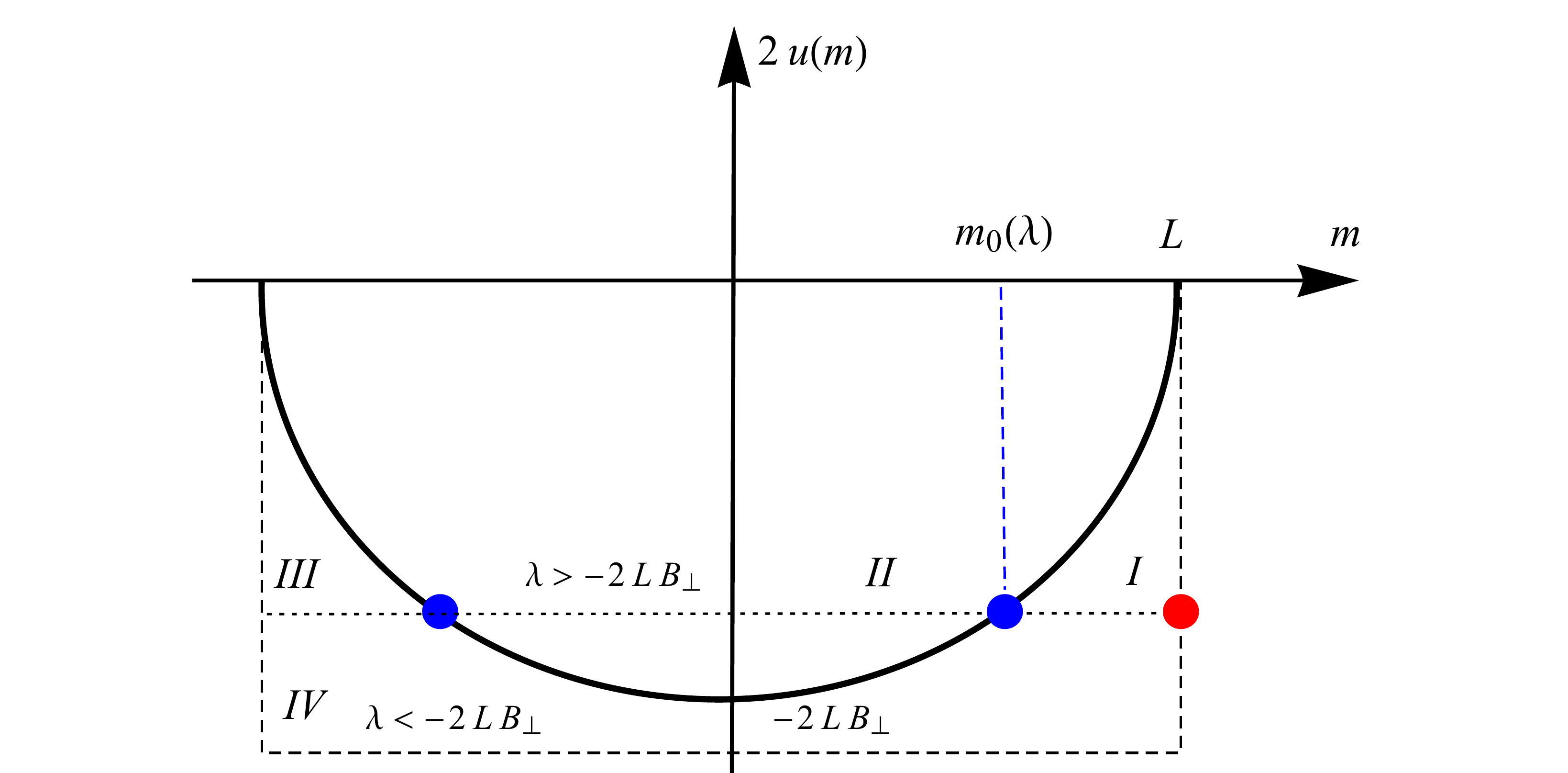}
\caption{ 
Effective potential for WKB analysis.
 }
\label{App:fig:EffectivePotential}
\end{figure}

Eq.~(\ref{eq:recur}) can be analyzed using WKB~\cite{DiscreteWKB} method resulting in,
\begin{gather}
 \mathbb{G}_{m,\frac{n}{2}} \propto  
 e^{i \int^m p(\lambda, q)\,dq},
 \end{gather}
 with quasiclassical momentum defined by,
 \begin{gather}
 -2 u(m) \cos p(\lambda, m) =\lambda,
 \end{gather}
 which can be explicitly inverted to obtain,
 \begin{gather}
 p(m, \lambda) =\arcsin \left(\sqrt{\frac{m_0^2-m^2}{L^2-m^2}} \right).
\end{gather}
Turning points, defined by the condition $p(m_0, \lambda)=0$,
\begin{gather}
m_0 = \frac{n}{2}\sqrt{1-a^{-2}},\;\;\; a=\frac{n B_{\perp }}{\left|\lambda\right|},
\end{gather}
separate classically forbidden and allowed regions of the Green function behavior, see Fig.~\ref{App:fig:EffectivePotential}, corresponding to exponential decay and oscillatory behavior respectively.

Within exponential accuracy the transverse field Green function takes the form, 
\begin{gather}
G(\lambda, d) \propto e^{- n S(a, d/n)} \sin \left( n \phi(a,d/n)\right),\label{eq:GvsS}
\end{gather}
where we used Stirling's asymptotic of the binomial coefficient,
\begin{gather}
S(a, \rho ) = \theta (a,\rho) -\frac{\rho}{2} \log(\rho) - \frac{(1-\rho)}{2}\log(1-\rho),
\end{gather}
where the phases $\theta (a,\rho)$ and $\phi (a,\rho)$ are determined by the large spin Green function $\mathbb{G}_{\frac{n}{2}-d,\frac{n}{2}}$.

In the case $a\equiv  \frac{n B_{\perp }}{\left|\lambda\right|}>1$ (regions $I, II, III$ in Fig.~\ref{App:fig:EffectivePotential}),
\begin{gather}
\theta (a,\rho)
 =
 \left\{
 \begin{array}{lcl}
  \int_{\frac{1}{2}-\rho }^{\frac{1}{2}} \mathrm{arcsinh} b(a,  \mu) \, d\mu,  &  (I) 
   \\ 
\theta_0,    & (II)
\\
 \int_{\frac{1}{2}-\rho }^{-\mu_0} \mathrm{arcsinh} b(a, \mu) \, d\mu +\theta_0,  & (III)
  \\
 \end{array}
 \right. \label{eq:Thetaa>1}
 \\
  \phi (a, \rho) = 
  \left\{
 \begin{array}{lcl}
  \frac{\pi}{2},  & (I) \\ 
\int_{\frac{1}{2}-\rho}^{\mu_0} \arcsin  \left| b(a,  \mu)\right| \, d\mu  +  \frac{\phi_0}{n},  & (II) \\
\frac{\pi}{2},  &  (III) \\
 \end{array} \right.
 \\
\theta_0 \equiv \int_{\mu_0}^{\frac{1}{2}} \mathrm{arcsinh} b(a,  \mu) \, d\mu.
\end{gather}

In the case $a\equiv  \frac{n B_{\perp }}{\left|\lambda\right|
 }<1$  (region $IV$),
\begin{gather}
\theta (a,\rho)= \int_{\frac{1}{2}-\rho}^{\frac{1}{2}} \mathrm{arcsinh} b(a,  \mu)\,d\mu, \;\;\; (IV), \label{eq:Thetaa<1}
\\
\phi(a, \rho) =\frac{\pi}{2}, \;\;\; (IV).
\end{gather}
In all of the above expressions,
\begin{gather}
b  (a, \mu ) = \sqrt{\frac{\mu^{2}-  \mu^2_0}{1/4  - \mu^{2}}}, \;\;\; \mu_0\equiv  \frac{m_0}{n}.
\end{gather}

\begin{figure}
\includegraphics[width=0.99\columnwidth]{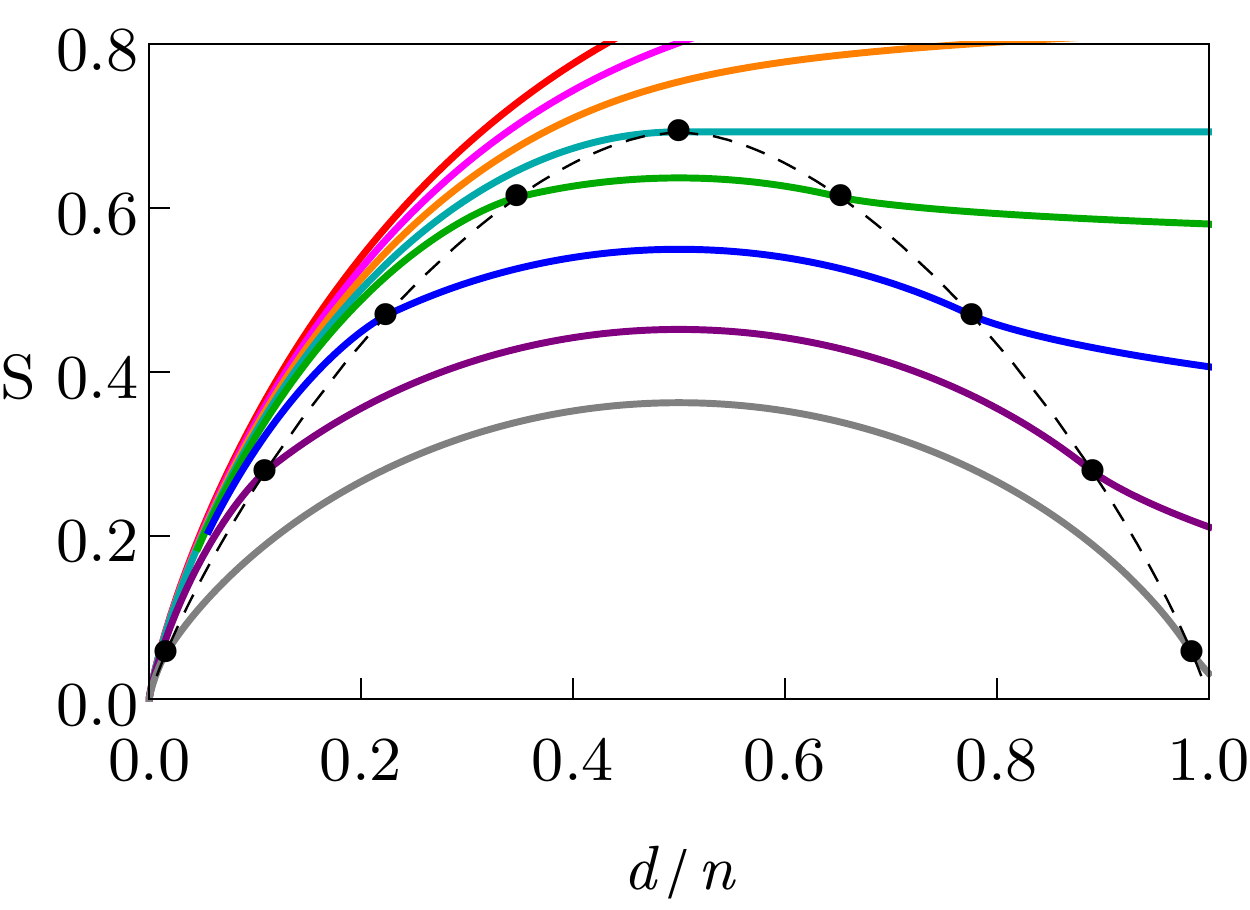}
\caption{ 
Exponent of the matrix element in Eq.~(\ref{eq:GvsS}) with $\theta(a,\rho)$ given in Eqs.~(\ref{eq:Thetaa>1},\ref{eq:Thetaa<1}). Colors correspond to different values of $0.9\leq  a  \leq 3$.
 }
\label{App:fig:MEExponent}
\end{figure}

Dependence of the exponent $S(\frac{n B_{\perp }}{\left|\lambda\right|}, d/n)$ on the Hamming distance $d/n$ is determined by the combination of the WKB exponent growing with $d/n$ and  QREM entropy which has a maximum at $d=\frac{n}{2}$, see Fig.~\ref{App:fig:MEExponent}. At weak transverse field  $\frac{n B_{\perp }}{\left|\lambda\right|}<1$, energy $-\left|\lambda\right|$ is below the ground state of the driver Hamiltonian $\hat{V}$. In this case the WKB exponent dominates and the tunneling matrix element decays exponentially for all $d/n$. At stronger transverse field $\frac{n B_{\perp }}{\left|\lambda\right|}>1$ the behavior depends on the region: $(I)$ matrix element decays exponentially;  $(II)$ the WKB exponent is real and results in oscillatory behavior of the Green function whereas the exponent is determined completely by the QREM entropy (see region below dashed line in Fig.~\ref{App:fig:MEExponent});  $(III)$ the growing WKB exponent competes with the decaying QREM entropy resulting in the exponent decaying with $d/n$. 
These two regimes are separated by the resonance $\frac{n B_{\perp }}{\left|\lambda\right|}=1$, $at d>n/2$ the WKB exactly compensates the QREM entropy and the exponent $S(a,d/n)$ is independent of Hamming distance and equals,
\begin{gather}
S\left(1,\rho\right) =- \frac{\rho}{2} \log \rho - \frac{(1-\rho)}{2} \log (1-\rho), 
\end{gather}
with $S\left(1,\frac{1}{2}\right) = \frac{1}{2} \log 2$. This resonance is exploited in the analog implementations of the Grover algorithm.

\begin{figure}[h]
\includegraphics[width=0.99\columnwidth]{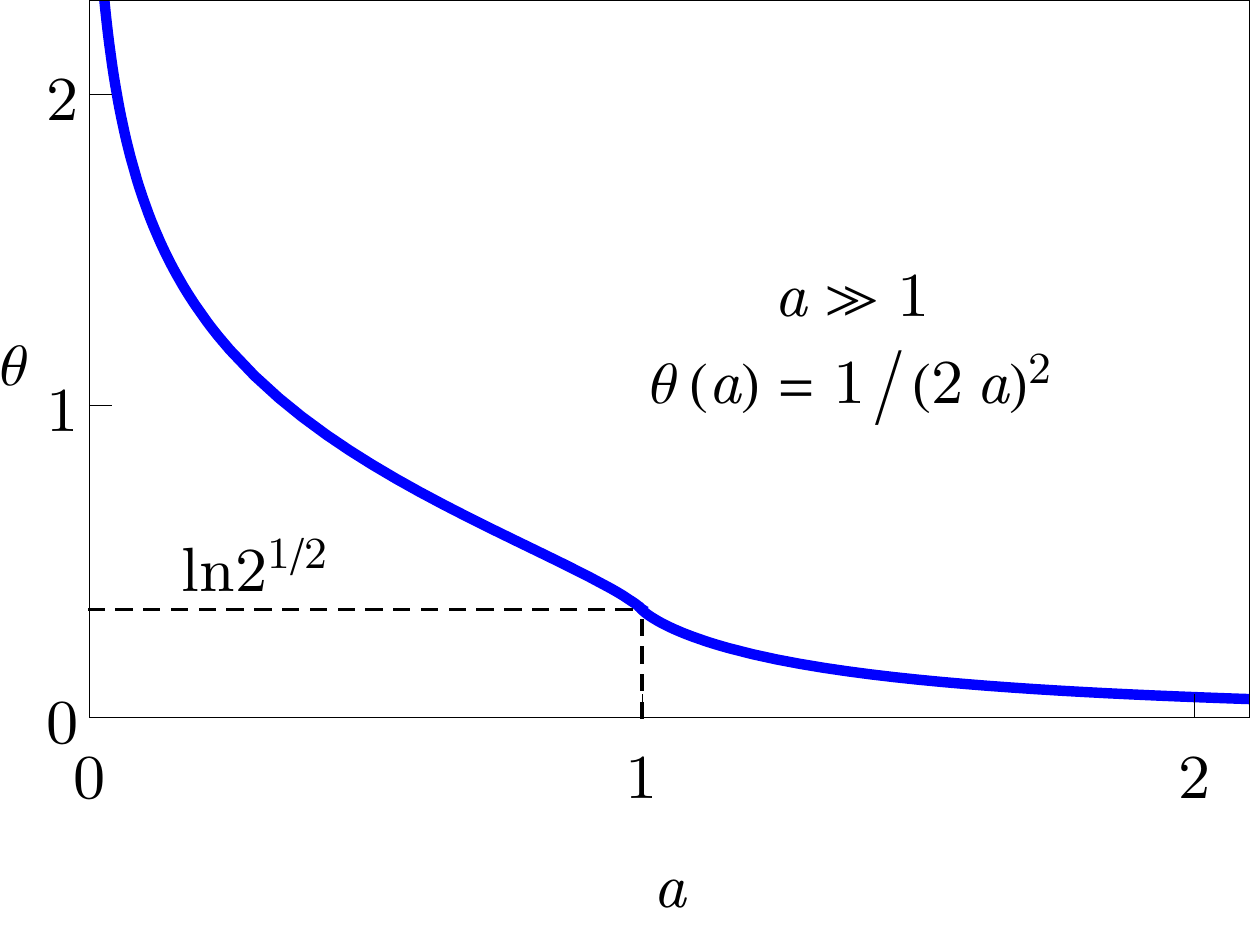}
\caption{ 
WKB exponent $\theta(a,\frac{1}{2})$ at $d=\frac{n}{2}$.
 }
\label{App:fig:MEatME}
\end{figure}

Fig.~\ref{App:fig:MEatME} shows matrix element at distance $d=\frac{n}{2}$ as a function of the relative transverse field parameter $a$.

\subsection{Weak disorder perturbation theory}

It is convenient to represent the series for the Green function $G(\lambda)$ in the form of diagrams shown in Fig.~\ref{fig:diagrams-SM}, where the solid lines correspond to the bare Green function $G_0(\lambda)$  and crosses to the "disorder potential" represented by the Hamiltonian $H_0$ which contains classical energies outside the downfolding subspace $\scS$ (energy strip $W$). Green function $\overline{G}$ is obtained by averaging with respect to realizations of disorder (shown with double line in Fig.~\ref{fig:diagrams-SM})
\begin{equation}
H_0=\sum_{j\in \bar{\scS}}\cE(s)\ket{s}\bra{s}
\end{equation}
 and taking into account that $\overline{\cE(s)}=0$ and $\overline{\cE^2(s)}=n \sigma^2/2$. 
 
 Introducing irreducible blocks in the diagramatic series $\Sigma(\lambda)$ (diagrams that cannot be split into disconnected pieces by cutting a single Green function line) and resumming the series in terms of these blocks Fig.~\ref{fig:diagrams-SM}(b) we obtain the Dyson equation for the average Green function $\overline{G}(\lambda)$, Fig.~\ref{fig:diagrams-SM}(c). The formal solution of the Dyson reads,
 \begin{gather}
 \overline{G}=(G_0^{-1}-\Sigma)^{-1}. \label{eq:DysonSolution}
 \end{gather}
 The self energy contains contribution of all possible scattering processes which can be represented as two types of blocks that we address separately below: (1) diagrams corresponding to multiple scattering on the same impurity; (2) diagrams corresponding to scattering off multiple impurities that include interference between them.
 
 The analysis of the diagramatic series requires understanding of the behavior of the matrix elements of the bare Green function and its tensor contractions with itself $\bra{s_1} G_0^p(\lambda)\ket{s_2}\equiv G_0^{(p)}(\lambda, d_{s_1 s_2})$ with Hamming distance $d_{s_1 s_2}$,
 \begin{widetext}
 \begin{gather}
 \bra{s} G_0^p(\lambda)\ket{s^\prime}\equiv G_0^{(p)}(\lambda, d_{s s^\prime}) =\frac{(-1)^{p-1}}{(p-1)!} \frac{d^{p-1}}{d\lambda^{p-1}}G_0(\lambda, d_{s s^\prime}), \nonumber \\
 G_0^{(\alpha)}(\lambda, d) = \frac{1}{2^n} \sum_{k=0}^{d}\sum_{K=0}^{n-d}(-1)^k 
 \binom{d}{k} \binom{n-d}{K} 
 \left(  
\frac{1}{\left(
\left(
 n-2(k+K) \right)B_\perp+\lambda
 \right)^\alpha} -\frac{1}{\lambda^\alpha}\right) \label{eq:ExactG0p}
 \end{gather}
  \end{widetext}
 
\begin{figure}
   \includegraphics[width=3.4in]{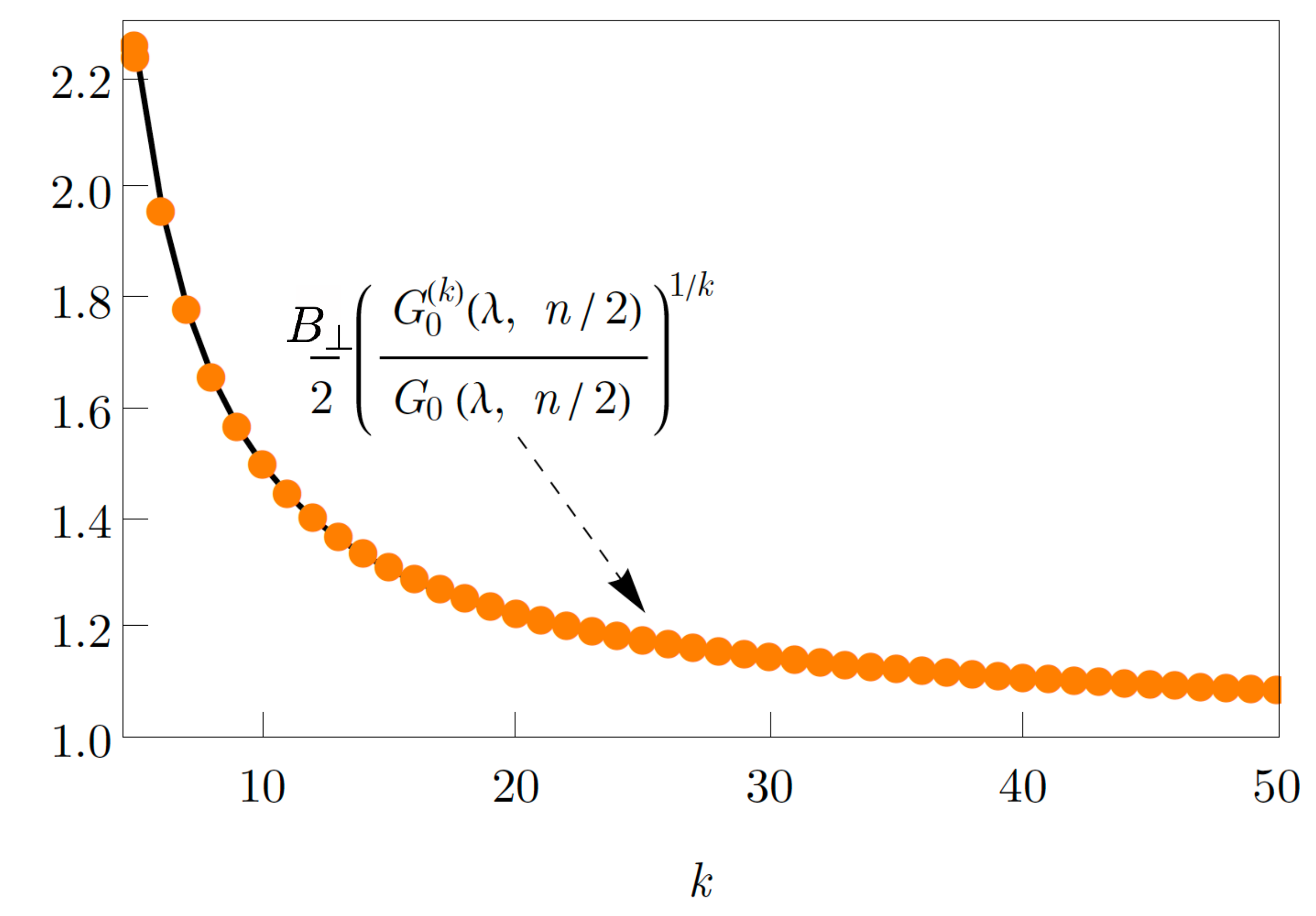}
  \caption{ Here we display a ratio of $G_0^{k)}(\lambda, d_{s_1 s_2})$ calculated using the exact expression Eq.~(\ref{eq:ExactG0p}) for $n=1000, B_\perp=100, \lambda=-B_\perp( n/2+1/2)$.
    \label{fig:G0vsd}}
\end{figure}

  We consider $p=1$ case first. For Hamming distances $d=\cO(1)$,
 \begin{gather}
 G_0(\lambda, d)  \approx  
 \frac{d!}{\lambda \left( -\lambda/ B_\perp\right)^{d}}=\cO\left(\frac{1}{n^{d+1}}\right), \;\; d=\cO(1). \label{eq:G0d1-SM}
 \end{gather}
 For $d=\cO(n)$ the matrix element is exponentially small,
 \begin{gather}
  G_0(\lambda, d)  \propto e^{-n S(B_\perp n/\lambda,d/n)}, \label{eq:WKBS-SM}
  \end{gather}
  where function $S(a,d/n)$ is depicted in the main text and explained in detail in the SM,  see Eq.~(\ref{eq:GvsS}). In higher order diagrams we encounter tensor contractions of the Green function. For small Hamming distances $d=\cO(1)$ we get,
 \begin{equation}
 G_0^{(p)}(\lambda, d) = \frac{(-1)^{p}(p+d-1)!}{(p-1)!} \frac{B_\perp^d}{\left| \lambda\right|^{d+p}} =\cO\left(\frac{1}{n^{p+d}}\right),
 \end{equation}
 We can now estimate for $d=\cO(1)$,
 \begin{gather}
 G_0^{(p)}(\lambda, d_{s_1 s_2})/G_0(\lambda, d_{s_1 s_2})=\cO(1/n^p).  \label{eq:G0pratio-SM}
 \end{gather}
 For $d=\cO(n)$ we have,
 \begin{gather}
G_0^{(p)}(\lambda, d) \propto \frac{1}{B_\perp^p} G_0(\lambda, d). \label{eq:G0pratioLarged}
 \end{gather}
 This relation could be obtained by comparing the first equation of Eq.~(\ref{eq:ExactG0p}) to the WKB expression Eq.~(\ref{eq:WKBS-SM}), also see Fig.~\ref{fig:G0vsd}. 
 
 We notice that for $d=\cO(1)$ the ratio of $ G_0^{(p)}(\lambda, d_{s_1 s_2})$ to $G_0(\lambda, d_{s_1 s_2})$ decreases exponentially with $p$ as $1/n^p$ from Eq.~(\ref{eq:G0pratio-SM}). This behavior changes drastically for $d=\cO(n)$ the $p$-order contraction of the Green function is of the same order as $ G_0(\lambda, d)$ itself (except for the factor $B_\perp^{-p}$ that can be small at large fields). 
   
 \subsection{ Multiple scatterings from the same impurity} 
 
\begin{figure}
   \includegraphics[width= 3.7in]{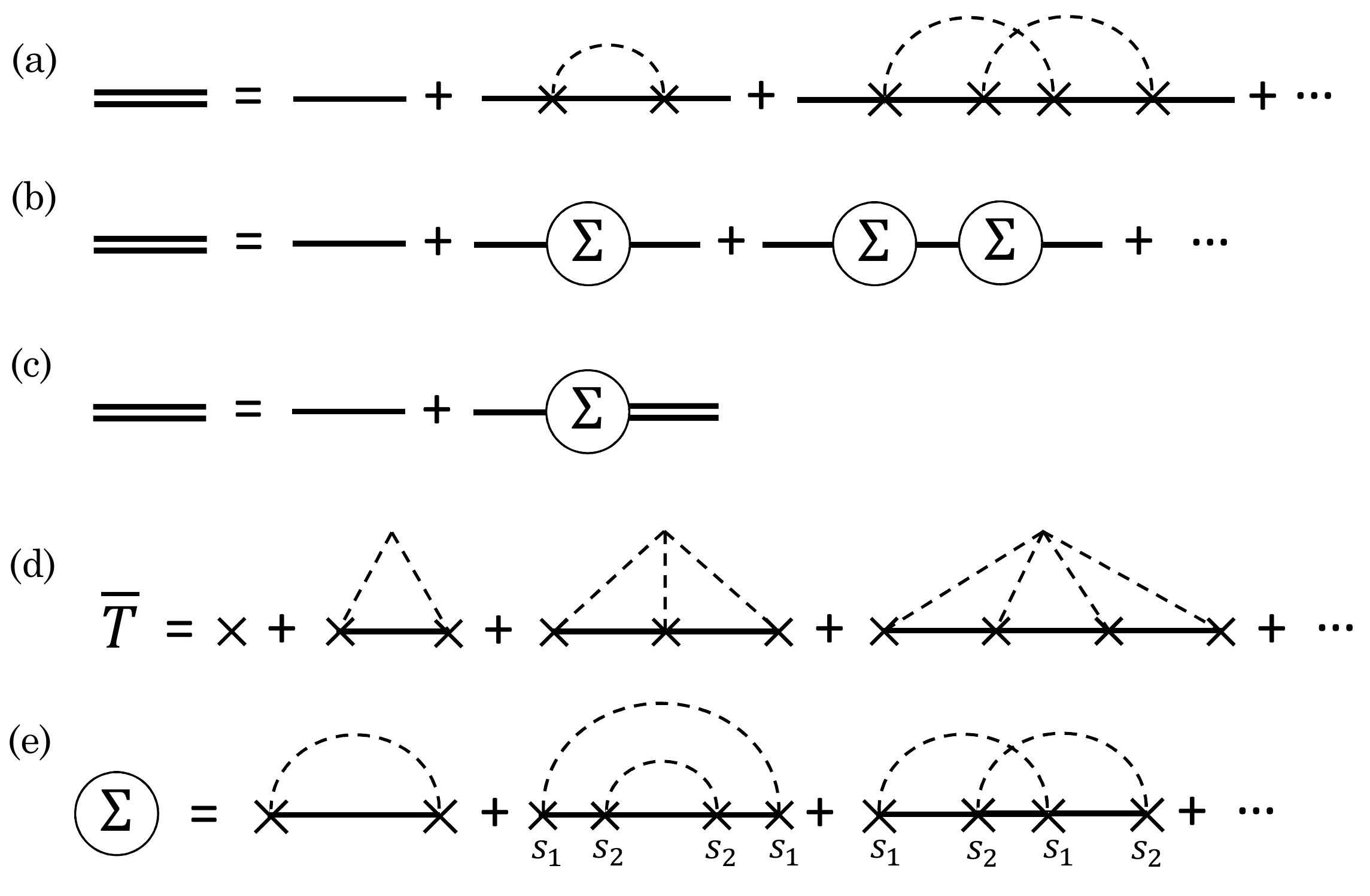}
  \caption{Diagrams 
    \label{fig:diagrams-SM}}
\end{figure}

 In the series of diagrams for $\Sigma$ multple scattering off the same impurity correspond to the "renormalization" of the disorder potential $\cE(s)$. Below we calculate this renormalization with the help of $T$-matrix.  We rewrite the Dyson series for $G(\lambda)$ before averaging in terms of $T$-matrix that accounts for scattering exactly,
\begin{gather}
G(\lambda)=G_0 + G_0 T G_0, \quad T=H_{0} \sum_{k=0}^{\infty} \left( G_0 H_{0} \right)^k.
\end{gather}
Effect of multiple scatterings on the  same impurity can be estimated  by independent impurities approximation.  We start considering a single impurity,
\begin{gather}
H_{0}^{(i)}=\cE(s_i)\ket{s_i}\bra{s_i},
\end{gather}
where $\cE(s_i)=\cO(\sqrt{n})$. In this simple case we can sum the perturbation series for the $T$-matrix exactly,
\begin{gather}
T_{i}=\frac{\cE_{i}}{1-\frac{\cE_{i}}{\lambda}}\ket{s_{i}}\bra{s_{i}},\quad
\overline{T}_{i} =  \frac{n \sigma^2}{2\lambda}\ket{s_{i}}\bra{s_{i}}+\cO\left(\frac{1}{n}\right).
\end{gather}
The average Green function for the case of a single impurity,
\begin{gather}
\overline{G}^{(i)}(\lambda)= G_0(\lambda) +  \frac{n \sigma^2}{2\lambda}G_0(\lambda)\ket{s_{i}}\bra{s_{i}} G_0(\lambda).
\end{gather}
Independent impurities approximation for the average Green function $\overline{G}^{\rm ind}$ corresponds to summing the second term over impurity locations resulting in the tensor contraction of two bare Green functions
\begin{gather}
\overline{G}^{\rm ind}(\lambda)= G_0(\lambda) +  \frac{n \sigma^2}{2\lambda}\left(G_0(\lambda)\right)^2+\cO\left(\frac{1}{n^3}\right)\;.
\end{gather}
Here 
 Note that this coincides exactly with the first non-vanishing correction in the disorder-averaged series depicted in Fig.~\ref{fig:diagrams-SM}(a). We conclude that  keeping terms corresponding to scattering on the same impurity at most twice in the disorder averaged series amounts to neglecting subleading corrections. This situation is analogous to the standard weak disorder perturbation theory where it is sufficient to keep track of only second order scattering amplitude and full $T$-matrix describing the exact scattering amplitude does not introduce qualitatively new terms in the perturbation series.  

\subsection {Scatterings from multiple impurities}

 In the following we account for interference of the trajectories scattered off multiple impurities. 
  The self-energy correction operator $\Sigma(\lambda)$ is given by the series in $H_0$ with the leading order corresponding to the first term in Fig.~\ref{fig:diagrams-SM}(e),
  \begin{gather}
  \Sigma^{(0)}=\frac{n\sigma^2}{2\cE} I=I\times \cO(1),
  \end{gather}
 already given in the main text, and $I$ is the identity matrix. The leading order expression for the Green function therefore reads,
 \begin{gather}
\overline{G}(\lambda) \approx G_0(\bar{\lambda}). \label{eq:Gbar0}
 \end{gather}
  The first corrections to the leading term in the self-energy $\Sigma^{(1)}_{A}+\Sigma^{(1)}_{B}$ are given by the second  and third diagrams in Fig.~\ref{fig:diagrams-SM}(e), respectively.
  
 We first consider $\Sigma^{(1)}_{A}$. Impurity lines (dashed lines in Fig.~\ref{fig:diagrams-SM}(a)) correspond to the constrains ${s_4=s_1\neq s_3=s_2}$ resulting in the following expression for the diagram,
\begin{gather}
\Sigma_{A}^{(1)}=I \left(\frac{ n\sigma^2}{2}\right)^2 G_0(\lambda,0)   G_0^{(2)}(\lambda,0)   =I \times \cO(1/n)\;. \nonumber
\end{gather}
Here again $I$ is an identity matrix, we used $\overline{ \cE^2 }= n\sigma^2/2$ and the asymptotical expression for $ \bra{s_1}  G_0^2(\lambda) \ket{s_1}$ given above. We remind the reader that $ G_0^{(2)}(\lambda,0) $ is the diagonal matrix element of the operator $\left( G_0(\lambda)\right)^2$, see the first Eq.~in~(\ref{eq:ExactG0p}).

The second contribution $\Sigma^{(1)}_{B}$, the third term in Fig.~\ref{fig:diagrams-SM}(e) corresponds to the constrain $s_3=s_1\neq s_4=s_2$ and equals
\begin{gather}
 \bra{s_1} \Sigma^{(1)}_{B} \ket{s_2}=\left(\frac{ n\sigma^2 }{2}\right)^2 \left( G_0(\lambda, d_{s_1 s_2}) \right)^3.  \nonumber
\end{gather}
In this expression we encounter for the first time non-zero off-diagonal matrix element of the operator $\Sigma(\lambda)$.  We now analyze the corrections from this term to $\overline{G}(\lambda)$ shown in Fig.~\ref{fig:diagrams-SM}(a),
\begin{widetext}
\begin{gather}
 \bra{s_i}G_0(\lambda)\Sigma^{(1)}_{B}(\lambda) G_0(\lambda)\ket{s_f} =\left(\frac{ n\sigma^2 }{2}\right)^2\sum_{s s^\prime} G_0(\lambda, d_{s_i s}) \left( G_0(\lambda, d_{s s^\prime}) \right)^3 G_0(\lambda, d_{s^\prime s_f})
 (1-\delta_{0,d_{ss^\prime}})
\end{gather}

Dominant contribution into the sum over $s,s^\prime$ is given by terms with $d_{s s^\prime}=1$, see Eq.~(\ref{eq:replace}) discussion in the next Section~\ref{sec:convolution}. Therefore we set in the above expression  $d_{s s^\prime}=1$ arriving at,
\begin{gather}
 \bra{s_i}G_0(\lambda)\Sigma^{(1)}_{B}(\lambda) G_0(\lambda)\ket{s_f} \approx \frac{\lambda}{B_\perp} \left(G_0(\lambda, 1)\right)^3( G^{(2)}(\lambda, d_{s s^\prime})-\frac{1}{\lambda} G_0(\lambda, d_{s s^\prime}))
\end{gather}
\end{widetext}
This quantity is smaller than $\overline{G}(\lambda)$ at least by the factor $\lambda\left(G_0(\lambda, 1)\right)^3=\cO(1/n^5)$. We conclude therefore that both self-energy corrections $\Sigma^{(1)}_{A}$ and $\Sigma^{(1)}_{B}$ correspond to at most $1/n$ correction to the respective matrix element of the Green function. 

Higher order self-energy corrections beyond those shown in Fig.~\ref{fig:diagrams-SM}(e) correspond to only topologically distinct diagrams such that their order cannot be reduced by replacing the bare Green function $G_0(\lambda)$ with its renormalized value $\overline{G}(\lambda)$. Such a replacement amounts to $1/n$ correction and therefore exceeds the accuracy of our approximation. Topologically distinct diagrams necessarily require crossings which correspond to traversing distance between distinct impurities $p\geq 2$ (more than two) times. As a result the structures of the form arise,
\begin{gather}
\sum_{s_2 s_3} O_{s_1 s_2} \left(G_0(\lambda, d_{s_2 s_3})\right)^{p} O^\prime_{s_3s_4}\approx \left(G_0(\lambda, 0)\right)^{p} \left(O O^\prime\right)_{s_1s_4},\nonumber
\end{gather} 
which is small as $ \left(G_0(\lambda, 0)\right)^{p}=\cO(n^{-p})$.  This above estimate follows from rapid decay with Hamming distance $d_{s_2 s_3}$ of  $ \left(G_0(\lambda, d_{s_2 s_3})\right)^{p}$. 

Due to the fact that the Green function matrix elements depend only on the Hamming distance and decay exponentially with it, 
the perturbation theory is analogous to the case of an under barrier scattering of a particle in finite dimension. We expect therefore that although the diagram series is complicated  the number of topologically distinct diagrams grows only polynomial with their order in the expansion parameter $1/n$. 
As a result the weak disorder effect can be described by the leading order diagram for $\Sigma (\lambda)$.

\subsection{Asymptotical properties of Green function tensor contractions}\label{sec:convolution}

Note that below we assume $d_{s_i s_f} =\cO(n)$. We calculate
\begin{gather}
K_{s_i s_f}^{(p)}=\sum_{s_1,s_2}G(\lambda, d_{s_i s_1}) \left(G(\lambda, d_{s_1 s_2})\right)^pG(\lambda, d_{s_2 s_f})
\end{gather}
we rewrite this expression in the following form
\begin{widetext}
\begin{gather}
K_{s_i s_f}^{(p)}=\sum_{d_1,d_2,d_3}^n G(\lambda, d_1) \left(G(\lambda, d_2)\right)^pG(\lambda, d_3) F_{s_i s_f}^{(3)}(d_1,d_2,d_3)\\
F_{s_i s_f}^{(3)}(d_1,d_2,d_3) =\sum_{s_1s_2} \delta_{d_1,d_{s_i s_1}}\delta_{d_2,d_{s_2 s_2}}
\delta_{d_3,d_{s_2 s_f}}
\end{gather}
in what follows we will need,
\begin{gather}
F_{s_i s_f}^{(2)}(d_1,d_2) \equiv F(d_1,d_2,d_{s_i s_f})=\sum_{s_1} \delta_{d_1, d_{s_i s_1}} 
\delta_{d_2, d_{s_1 s_f}}.
\end{gather}
One can show that,
\begin{gather}
F(d_1,d_2,d_3)=\binom{d_3}{\frac{1}{2}(d_3-d_1+d_2)} \binom{n-d_3}{\frac{1}{2}(d_1+d_2-d_3)}
\end{gather}
We not the reduction formula for kernels $F$,
\begin{gather}
F_{s_i s_f}^{(3)}(d_1,d_2,d_3) =\sum_{d=0}^{n} F(d_1,d_2,d)F(d,d_3,d_{s_i s_f})
\end{gather}
We can rewrite an arbitrary convolution of the Green functions in terms of convolutions of Green functions in Hamming distance only. For example,
\begin{gather}
\sum_{s_1,s_2}\left(G(\lambda, d_{s_i s_1})\right)^q \left(G(\lambda, d_{s_1 s_2})\right)^b \left(G(\lambda, d_{s_2 s_f})\right)^r \nonumber \\
=
\sum_{d,d_1,d_2,d_3=0}^{n}   \left(G(\lambda, d_1)\right)^q \left(G(\lambda, d_1)\right)^b \left(G(\lambda, d_3)\right)^r F(d_1, d_2, d) F(d, d_3, d_{s_i,s_f})
\end{gather}
The above equation allows to express arbitrary convolutions of Green functions in terms of convolutions of their $F$-transforms,
\begin{gather}
A(\lambda, d_2, d)=\sum_{m=0}^{n} G(\lambda, m) F(m, d_2, d)
\end{gather}
For example,
\begin{gather}
\sum_{s_1,s_2}G(\lambda, d_{s_i s_1}) \left(G(\lambda, d_{s_1 s_2})\right)^p G(\lambda, d_{s_2 s_f}) = \sum_{d_2} \left( G(\lambda, d_2)\right)^p \left( A^2 \right)_{d_2 d_{s_i s_f}} 
\approx
\left( G(\lambda,0)\right)^p \left( A^2 \right)_{0 d_{s_i s_f}} 
\end{gather}
The latter approximate equality is result of the fact that $ \left(G(\lambda, d)\right)^p$ where $p>1$ decreases with Hamming distance $d$ faster than the factor $ \left( A^2 \right)_{d d^\prime} $. 
It is easy to show that
\begin{gather}
\left( A^2 \right)_{0 d_{s_i s_f}}  = G^{(2)}(\lambda, d_{s_i s_f})
\end{gather}
 Therefore the convolution 
 \begin{gather}
 \sum_{s_1,s_2}G(\lambda, d_{s_i s_1}) \left(G(\lambda, d_{s_1 s_2})\right)^p G(\lambda, d_{s_2 s_f})  
 =
 \frac{n^2}{4} \left( G(\lambda, 0)\right)^p G^{(2)}(\lambda, d_{s_i s_f})\sim \frac{1}{n^p} G(\lambda,  d_{s_i s_f}) 
 \end{gather}
 \begin{figure}
   \includegraphics[width= \columnwidth]{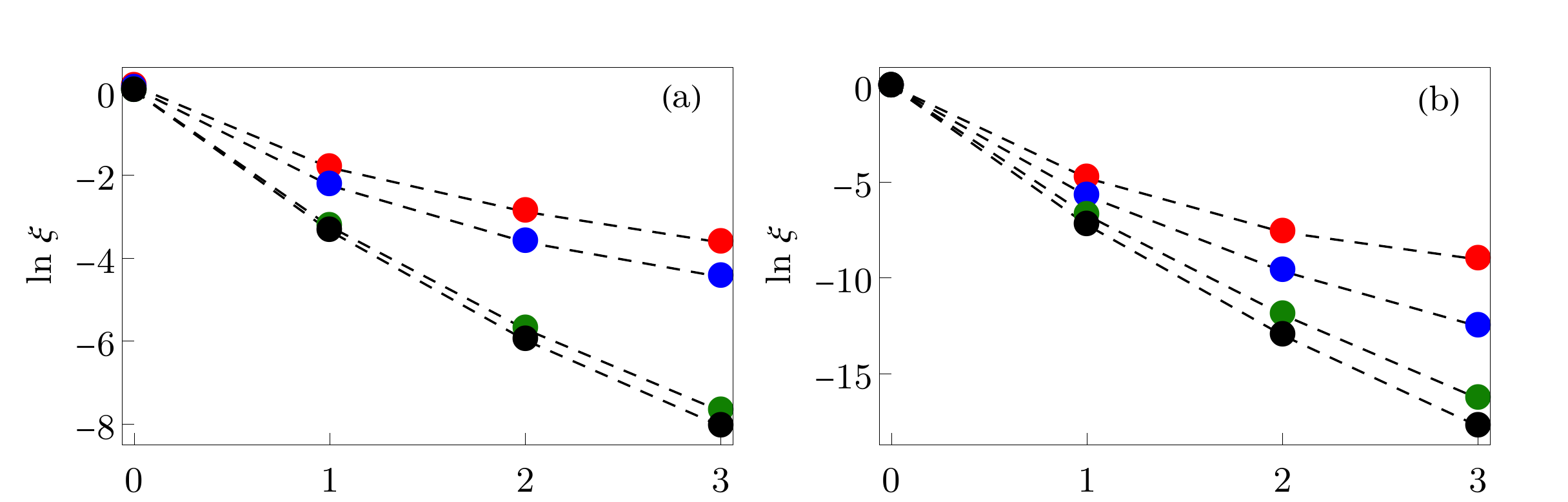}
  \caption{The plot of the function $\ln \xi_k(\lambda)$  vs. $k$, see Eq.~(\ref{eq:xik}). Dashed line are guides to the eye. Different colors correspond to different numbers of spins $n$: 40 (red), 60 (blue), 70 (green), 100 (black). Parameters used: $B_\perp=2, d_{s_i s_f}, \lambda=-B_\perp (n/2+1/2)$.
    \label{fig:contraction}}
\end{figure}

The above property  is a manifestation of a general rule, a general multiple contraction of the Green function can be approximated in the leading order in $1/n$,
 \begin{gather}
\sum_{s,s^\prime}O_{s_i s}\left( G_0(\lambda, d_{s s^\prime})\right)^pO^\prime_{s^\prime s_f}\approx \sum_{s,s^\prime}O_{s_i s}\delta_{0,d_{s s^\prime}}\left( G_0(\lambda, 0)\right)^pO^\prime_{s^\prime s_f}\label{eq:replace}
 \end{gather}
 We note that including in the convolution terms with $d_{s s^\prime} =1,2,$ etc. will give rise to corrections $\cO(1/n), \cO(1/n^2),$ etc., respectively.
 To illustrate the above we plot the following quantity,
\begin{gather}
\xi_k(\lambda) = \frac{\sum_{s s^\prime} \theta(d_{s s^\prime}-k) G_0(\lambda, d_{s_i s})  \left(G_0((\lambda, d_{s_i s}) \right)^3  G_0(\lambda, d_{s^\prime s_f}) }
{
\left(G_0(\lambda, 0) \right)^3 G^{(2)}(\lambda, d_{s_i s_f})
} \label{eq:xik}
\end{gather}
One can see from Fig.~\ref{fig:contraction} that asymptotical behavior $\xi_k(\lambda) \propto 1/n^k$ is reached at large $n$  and the maximum is at $k=1$ which corresponds to the approximation  in Eq.~(\ref{eq:replace}).
 \end{widetext}

\section{Partition function}

Coexistence of asymptotically orthogonal bands of $z$- and $x$- type states implies that  the partition function splits into two contributions, 
\begin{gather}
\mathcal{Z}=\mathcal{Z}_z+\mathcal{Z}_x,
\end{gather} 
which can be calculated analyzing and  separately for  each set of states. $z$-bands are described in detail in the main text.  Here we focus on $x$-bands.

\subsection{$x$-type states} 
These states conserve total magnetization along $x$-axis and can be conveniently written in the respective eigenbasis 
\begin{equation}
S^x\ket{x^i}=(n - 2 m)\ket{x^i},\quad \ket{x^i}=\bigotimes_{k=1}^{n}\ket{x^i_k}
\end{equation}
\begin{equation}
\sigma^x_k \ket{x^i_k}=(1-2x^i_k)\ket{x^i_k}, \quad x^i_k=0,1
\end{equation}
where   $S^x\equiv \left(1/n\right) \sum_{k=1}^n\sigma^x_k$ and $m=0,1,\ldots,n$. The eigenstates of $S^x$ are mixed by the matrix elements of $H_{REM}$,
\begin{gather}
\bra{x} H_{cl}\ket{x'}=\frac{1}{2^n}\sum_{i=0}^{2^n-1} (-1)^{\sum_k(x_k+x'_{k})s_k^i} \cE_i,\\
\overline{\left(\bra{x} H_{cl}\ket{x'}\right)^2}=\frac{1}{2^{2n}}\sum_{i=0}^{2^n-1} \langle \cE_i^2 \rangle=\frac{n}{2^{n+1}}, \label{HxxVar}
\end{gather} 
where $\overline{(...)}$ stands for averaging over different realizations of $\cE_i, i=0,...,2^n-1$. The diagonal part $\bra{x} H_{cl}\ket{x}=\frac{1}{2^n}\sum_{i=0}^{2^n-1} \cE_i$ is uniform. 

Leading order description of the $x$-states requires taking into account virtual transition to the band of states at vanishing energy density $\epsilon  \rightarrow 0$. We can use an analog of downfolding procedure described in the main text which we adapt here for the $x$-basis. We consider a set of $x$-basis states characterized by the same magnetization $n-2m$, which will be the downfolding subspace $\scS_x$. The non-linear eigenvalue problem arising from the downfolding procedure reads,
\begin{gather}
\left(H^{(m,m)}+\zeta \right)\Psi =\lambda  \Psi,\\
\zeta_{x,x'}\equiv \sum _{y, y'} H_{x,y}^{(m,m')}  \mathcal{G}_{y,y'}\left(\lambda \right) H_{y',x'}^{(m'',m)}.
\end{gather}
Here $\mathcal{G}_{y,y'}\left(\lambda \right) $ is the Green function defined outside the downfolding subspace $\scS_x$. The weak disorder perturbation theory described in the main text and explained in detail in the above Sections of the SM shows that the effect of random energies $H_{cl}$ on the Green function of a low energy state $\lambda =\cO(n)$ is to weakly renormalize the energy ${\lambda \rightarrow \bar{\lambda} \approx\lambda - n\sigma^2/\lambda}$. Therefore we can approximate
\begin{gather}
 \mathcal{G}_{x,x'}^0\left(\lambda \right) \approx \delta _{x,x'}\frac{\ket{x}\bra{x}}{\lambda-2 B_{\perp}(n-2m)}
 \end{gather}

\begin{gather}
\overline{\zeta}_{x,x'}\approx \frac{1}{2 (1-2 m/n ) B_{\perp}}, \\
\overline{\zeta _{x,x'}^2}-\overline{\zeta _{x,x'}}{}^2=\frac{1}{\left(1-\frac{2 m}{n}\right)^2B_{\perp}^2}\frac{2}{2^n}.
\end{gather}

Off-diagonal terms are Guassian random variables with variance given by Eq.~(\ref{HxxVar}). 
In the leading order expansion in $\hat{H}$ mixing of the bands with different projections on the $x$-axis, $m'\neq m$, can be neglected and therefore the set of  $\binom{n}{m}$ states with fixed $x$ magnetization $m$ can be approximated by a Gaussian Orthogonal Ensemble (GOE) of random matrix theory. The Wigner's semicircle law for GOE predicts the many-body band width 
\begin{gather}
\Gamma_m^x \sim \sqrt{\frac{n}{2}\binom{n}{m}}2^{-n/2},
\end{gather}
which is exponentially small up to $m -  n/2 \sim \mathcal{O}  (n^{1/2}) $.

\subsection{Cumulative density of states}

\begin{figure}[t]
\includegraphics[width=0.99\columnwidth]{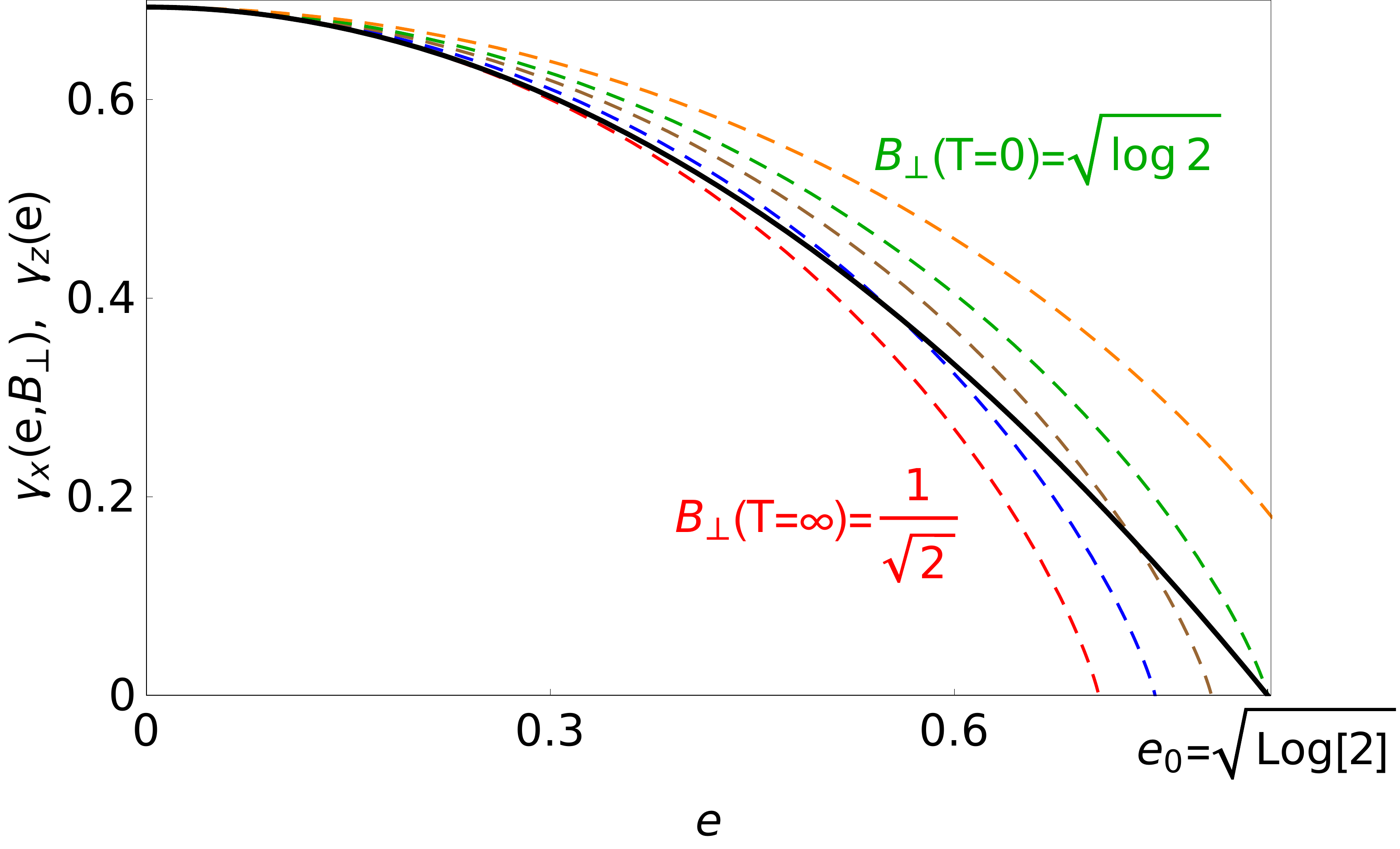}
\caption{ 
Partial entropy of $z$ and $x$ states as a function of energy per spin for different magnetic fields.
 }
\label{fig:PartialEntropies}
\end{figure}
We compare cumulative density of states of $x$ and $z$ type states. For the former the sum over  the number of states in $x$-state minibands labeled by $x$-magnetization $n-2m$ give the cumulative density,
\begin{gather}
C_x(e, \B)=\sum_{m=0}^{m_{\ast}(e)} \binom{n}{m},
\end{gather}
where $m_{\ast}(e)=n/2-en/(2\B)$ is the maximum included miniband index. The partial entropy of  the $x$-states ${\gamma_x = \lim_{n\rightarrow\infty} \frac{1}{n}\ln C_x(e, \B)}$, reads,
\begin{gather}
\gamma_x = 
 \ln 2 -\frac{1}{2} \ln \left( 1-\frac{e^2}{B_\perp^2}\right) -\frac{1}{2}\frac{e}{B_\perp} \ln\frac{  1-\frac{e}{\B}}{1+\frac{e}{\B}}.
\end{gather}
Partial entropy of $z$-states is that of REM,
\begin{gather}
\gamma_z = \ln 2 -e^2,
\end{gather}
shown as solid black line in Fig.~\ref{fig:PartialEntropies}. 
The phase transition corresponds to the exponents $\gamma_x = \lim_{n\rightarrow\infty} \frac{1}{n}\ln C_x(e, \B)$ and $\gamma_x$ being equal to each other,
\begin{gather}
\Delta \gamma =\gamma_z(e) -\gamma_x(e,\B) =0,
\end{gather}
which correspond to the crossing of the solid black line in Fig.~\ref{fig:PartialEntropies} with dashed lines showing $\gamma_x(e, \B)$. 
\begin{figure}[t]
\includegraphics[width=0.99\columnwidth]{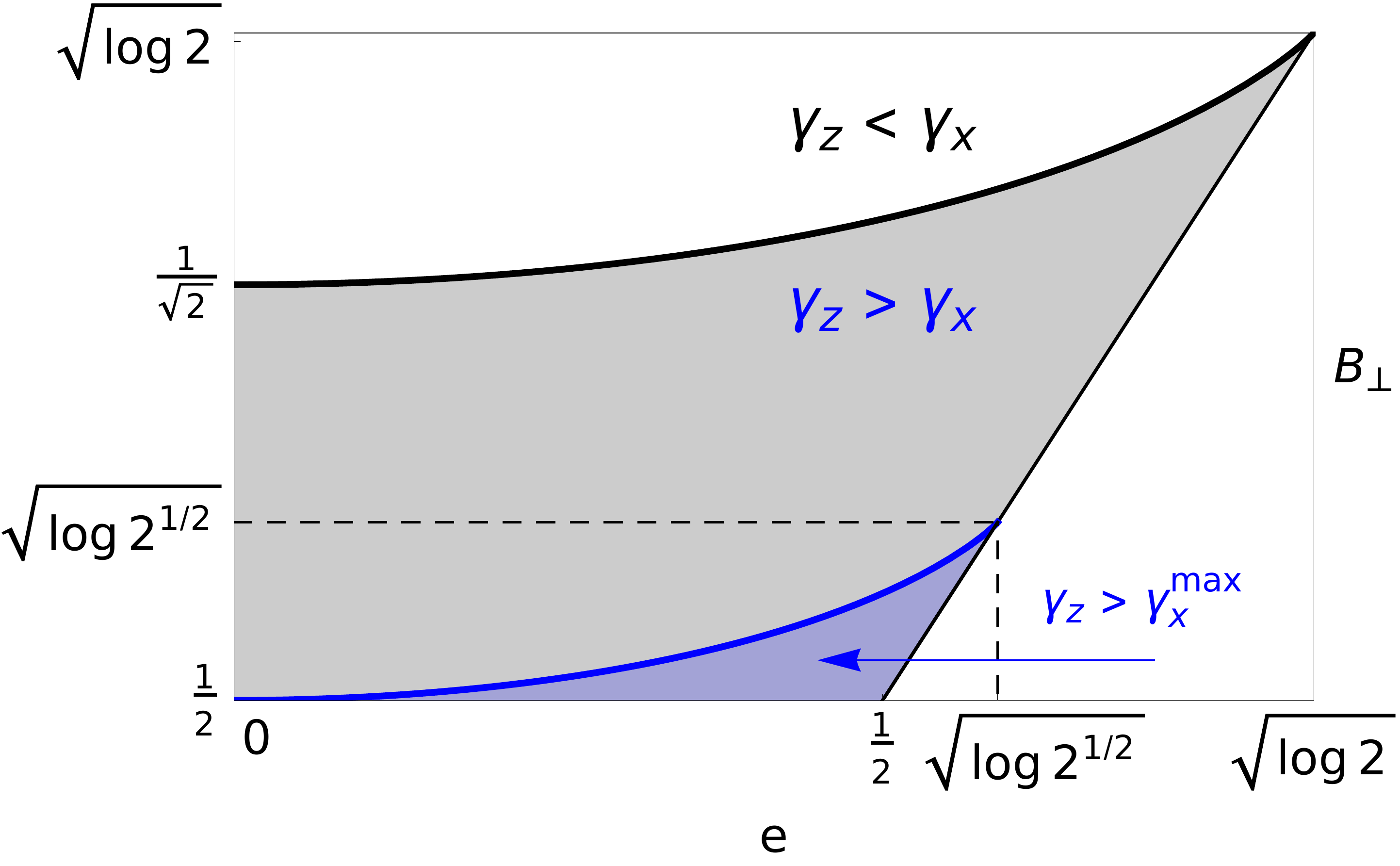}
\caption{ 
Statistical phase diagram of the QREM eigenstates on the $(e,  \B)$ plane. Gray shaded region corresponds to the phase where partial entropy $z$-states is larger than that of $x$-states and the spectrum consists of nearly uniform density of $z$-states with sharp and narrow peaks of the desnity of $x$ states around eigenvalue $\lambda=-2\B m$.  In the region shaded blue $z$-state density is larger than the peak density of $x$-states. See text for details.   
 }
\label{fig:DoS-SM}
\end{figure}
The resulting phase transition line is depicted as the solid line in Fig.~\ref{fig:DoS-SM}.  At energy densities $e\ll1$,
\begin{gather}
\Delta \gamma \approx e^2 \frac{1/2-B_\perp^2}{B_\perp^2},
\end{gather}
giving the phase transition point $B_\perp=1/\sqrt{2}$.  $\Delta \gamma=0$ line (black solid line in Fig.~\ref{fig:DoS-SM}) separates the region in the spectrum where there are exponentially more $x$-states $\gamma_x(e, B_\perp)>\gamma_z(e)$, from the region where there are exponentially more $z$-states $\gamma_x(e, B_\perp)<\gamma_z(e)$ shown as gray area in the Fig.~\ref{fig:DoS-SM}.   Within the latter region the density of states of $x$-states consists of exponentially narrow peaks corresponding to the NEE minibands separated by wide energy gaps $\sim  2\B$ filled with nearly uniform density of $z$-type states. Blue solid line in Fig.~\ref{fig:DoS-SM},  
\begin{gather}
\Delta \gamma_{max}=-e^2 +\frac{1}{4}\ln\left(1-\frac{e^2}{B_\perp^2}\right)+\frac{1}{4} \ln\frac{ 1+\frac{e}{\B}}{1-\frac{e}{\B}},
\end{gather}
separates the region (gray region in the figure) where the maximum of the density of $x$-states is larger than the density of $z$-states despite the latter dominating in the cumulative density of states.  In the region shaded with transparent blue density of $z$-states is larger than the maximum of the density of $x$-states. This level structure is illustrated by the cartoon in the main text Fig.~1
(d).

\subsection{Finite temperature partition function}

In this Section we use the statistics of the eigenstates to compare to the earlier replica analysis of the QREM partition function with static approximation.

\subsubsection{$x$-state partition function}

\begin{figure}[t]
\includegraphics[width=0.8\columnwidth]{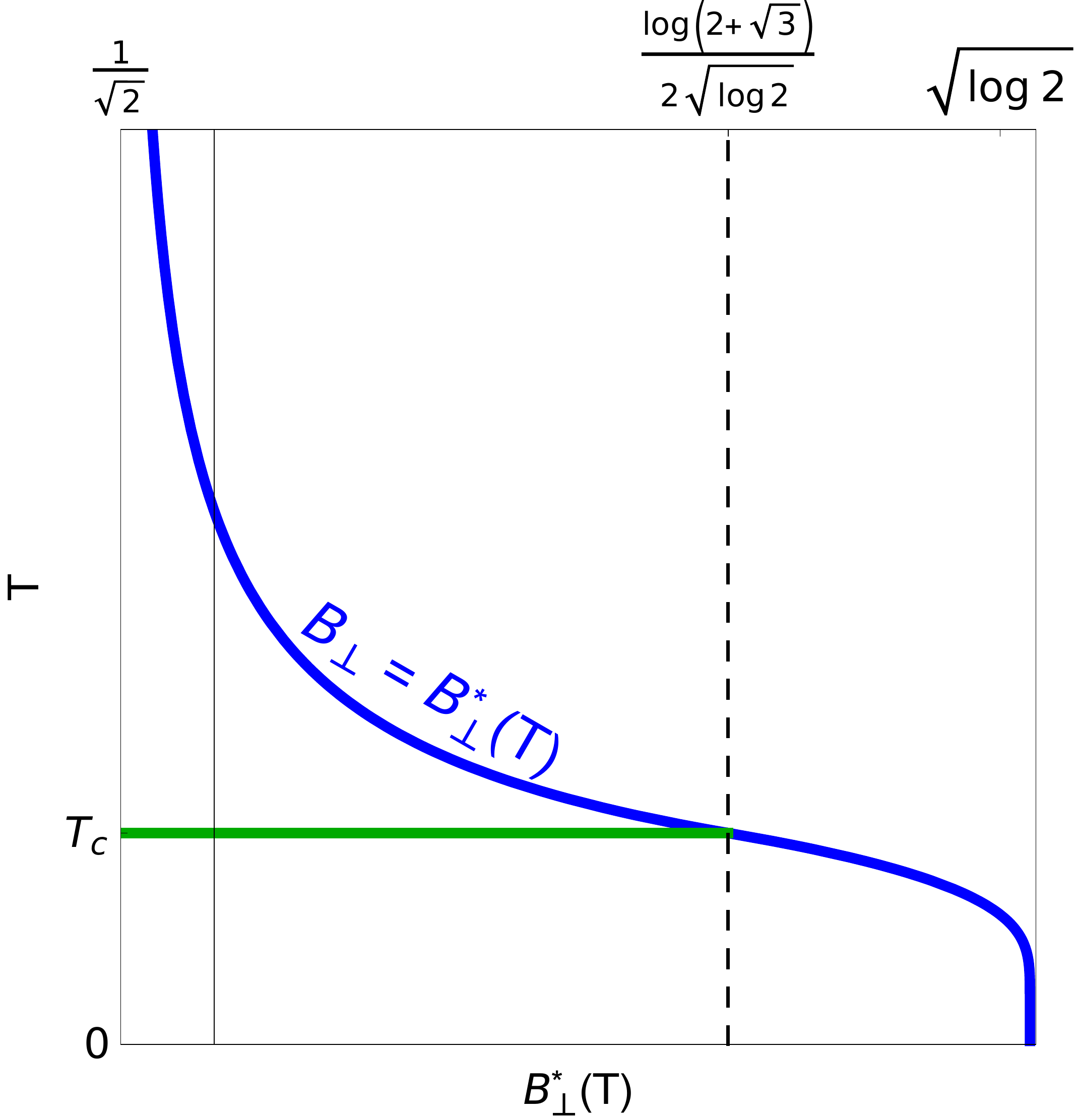}
\caption{ 
Classical to quantum paramagnet phase transition line.
 }
\label{fig:PhaseTransitionLineTe-SM}
\end{figure}

Partition of function of $x$-states consists of two parts: that of the free spin in transverse field $\B$, and a correction due to Gaussian Orthogonal ensemble statistics $\Delta Z_x(m,\beta )$,
\begin{gather}
 Z_x (\beta)=2^n \left[ \cosh B_{\perp}\beta \right]^n+\Delta Z_x(m,\beta ), \\
\Delta Z_x(m,\beta )=\sum _{m=0}^n e^{-\beta  E_m} \binom{n}{m}
\left(_0F_1\left(2,\Xi\right)-1\right),\\
\Xi\equiv \frac{1}{2} 2^{-n} \left(\beta ^2 n\right) \binom{n}{m}.
\end{gather}
The corresponding free energy reads,
\begin{gather}
f=-\frac{1}{n}\ln\left(\beta  Z_x-\Delta Z_x(m,\beta )\right).
\end{gather}
The above expression can be expanded in the low temperature limit $\beta\gg1$ as follows,
\begin{gather}
f\approx-\frac{1}{2} \beta ^2 B^2-\ln (2),
\end{gather}
with the correction estimated as,
\begin{gather}
-\frac{1}{n}\ln \Delta Z_x(m,\beta ) = 1/6 B^4 \beta^4 -\ln[2].
\end{gather}
Comparison of the two contributions to the partition function,
\begin{gather}
\frac{\Delta Z_x(m,\beta )}{\beta  Z_x-\Delta Z_x(m,\beta )}=\mathcal{O}\left( 2^{-\alpha  n}\right),
\end{gather}
shows that the correction can be safely neglected.

\subsubsection{$z$-state partition function}

Free energy of QREM $z$-states is described by the classical expression,
\begin{gather}
f_z(\beta) =\left\{
\begin{array}{ll}
-\frac{1}{4} \beta -\beta^{-1} \ln 2, & \beta\leq \beta_c \\
-\sqrt{\ln 2},  & \beta>\beta_c
\end{array}
\right.,
\end{gather}
where $\beta_c=2\sqrt{\ln2}$.

\subsubsection{Phase transition line}
Is determined from $f_{\text{Gibbs}}^x=f_{\text{Gibbs}}^z$ reads,
\begin{gather}
B_\perp^{\ast} =
\left\{
\begin{array}{ll}
\frac{1}{\beta}  \rm{arccosh}\left( \frac{1}{2}  \exp\left(\frac{\beta^2}{4} + \ln2 \right)\right), & \beta\leq \beta_c \\
\frac{1}{\beta}  \rm{arccosh}\left( \frac{1}{2}  \exp\left(\beta\sqrt{ \ln2} \right)\right),  & \beta>\beta_c
\end{array}
\right.,
\end{gather}
shown as the solid line in Fig.~\ref{fig:PhaseTransitionLineTe-SM}.

\end{document}